\documentclass[aps, prb, twocolumn, a4paper, superscriptaddress]{revtex4-2}

\usepackage{graphicx} 
\usepackage{dcolumn} 
\usepackage{amsmath, bm} 
\usepackage{hyperref} 
\usepackage{color} 
\usepackage{gensymb} 

\makeatletter
\g@addto@macro\bfseries{\boldmath}
\makeatother

\def\musr{$\mu^+$SR}
\def\gavs{GaV$_4$S$_8$}
\def\neel{N{\'e}el}

\newcommand{\vect}[1]{\ensuremath{\bm{#1}}}
\newcommand*{\hham}{\hat{\mathcal{H}}}

\begin{document}

\title{Magnetism in the N\'{e}el skyrmion host \texorpdfstring{\gavs}{GaV4S8}\ under pressure}

\author{T.~J.~Hicken}
\altaffiliation{Current address: Department of Physics, Royal Holloway, University of London, Egham, TW20 0EX, United Kingdom}
\affiliation{Department of Physics, Centre for Materials Physics, Durham University, Durham, DH1 3LE, United Kingdom}
\author{M.~N.~Wilson}
\affiliation{Department of Physics, Centre for Materials Physics, Durham University, Durham, DH1 3LE, United Kingdom}
\author{S.~J.~R.~Holt}
\affiliation{Department of Physics, University of Warwick, Coventry, CV4 7AL, United Kingdom}
\author{R.~Khassanov}
\affiliation{Laboratory for Muon Spin Spectroscopy, Paul Scherrer Institute, 5232 Villigen PSI, Switzerland}
\author{M.~R.~Lees}
\affiliation{Department of Physics, University of Warwick, Coventry, CV4 7AL, United Kingdom}
\author{R.~Gupta}
\affiliation{Laboratory for Muon Spin Spectroscopy, Paul Scherrer Institute, 5232 Villigen PSI, Switzerland}
\author{D.~Das}
\affiliation{Laboratory for Muon Spin Spectroscopy, Paul Scherrer Institute, 5232 Villigen PSI, Switzerland}
\author{G.~Balakrishnan}
\affiliation{Department of Physics, University of Warwick, Coventry, CV4 7AL, United Kingdom}
\author{T.~Lancaster}
\affiliation{Department of Physics, Centre for Materials Physics, Durham University, Durham, DH1 3LE, United Kingdom}

\date{\today}

\begin{abstract}
	We present magnetization and muon-spin spectroscopy measurements of \neel\ skyrmion-host \gavs\ under the application of hydrostatic pressures up to $P=2.29$~GPa.
	Our results suggest that the magnetic phase diagram is altered with pressure via a reduction in the crossover temperature from the cycloidal (C) to ferromagnetic-like state with increasing $P$,  such that, by 2.29~GPa, the C state appears to persist down to the lowest measured temperatures.
	With the aid of micromagnetic simulations, we propose that the driving mechanism behind this change is a reduction in the magnetic anisotropy of the system, and suggest that this could lead to an increase in stability of the skyrmion lattice.
\end{abstract}

\maketitle

\section{Introduction}
The lacunar spinel \gavs~\cite{brasen1975magnetic} is one of very few examples of a material that hosts a magnetic \neel\ skyrmion lattice (SkL)~\cite{lancaster2019skyrmions} throughout the bulk~\cite{kezsmarki2015neel}.
The SkL, along with the cycloidal (C) phase, stabilized in different parts of the applied field-temperature ($B$--$T$) phase diagram, occurs due to a competition between magnetic exchange and the Dzyaloshinskii-Moriya interaction (DMI).
The latter arises owing to the the polar rhombohedral structure ($R3m$) that \gavs\ adopts below a structural phase transition at 42~K~\cite{holt2020structure}, generally attributed to a Jahn-Teller distortion.
Above this temperature the material possesses a  cubic $F\bar{4}3m$ structure.
Also important in this material is easy-axis anisotropy~\cite{ehlers2016exchange}, whose direction aligns with the rhombohedral distortion.
Competition between the three interactions leads to a rich magnetic phase diagram, with dependence on the alignment between the applied magnetic field and crystallographic axes~\cite{kezsmarki2015neel}.
In zero field the ground state comprises a complex ferromagnetic-like (FM*) configuration, where the spins on the V atoms in each V$_{4}$ cluster combine to form effective spin-1/2 units that align ferromagnetically~\cite{holt2021investigation}.
On increasing temperature, C ordering then takes place in the region $5~\lesssim T<T_{\mathrm{c}}\simeq13$~K. 
Under the application of an applied field, in polycrystalline samples of \gavs, the SkL phase is stabilized at $10\lesssim T\lesssim 13$~K, $40\lesssim \mu_0H_\text{ext}\lesssim 100$~mT~\cite{hicken2020magnetism}.

The application of hydrostatic pressure on \gavs\ has been studied previously at room temperature~\cite{wang2021semiconducting}.
Above 35 GPa there is a pressure induced structural phase transition that results in the high-$T$ cubic phase becoming orthorhombic ($Imm2$), alongside a semiconductor to metallic transition and changes in the optical properties of the material.
At lower pressures, the changes in crystal structure are more modest, with a gradual reduction in the lattice parameter on increasing pressure.
Although there has been some effort to construct a magnetic $T$--$P$ phase diagram of \gavs~\cite{mokdad2019structural}, the transitions between the magnetically ordered states have not been thoroughly investigated.

We have previously studied the magnetism in the GaV$_4$S$_{8-y}$Se$_y$ series using magnetometry and muon-spin spectroscopy (\musr)~\cite{franke2018magnetic,hicken2020magnetism}.
These techniques are sensitive to subtle changes in the magnetism of this series, revealing, for example, that the transition from  FM* to C states is a crossover, rather than an abrupt phase transition~\cite{white2018direct,clements2020robust}.
This process is likely to depend sensitively on the crystalline anisotropy in the system~\cite{izyumov1984modulated}, which has been shown to decrease as $T$ increases~\cite{ehlers2016exchange}.
Here we extend our investigation of \gavs\ to probe the magnetic behavior under the application of hydrostatic pressure up to $P=2.29$~GPa.
With the aid of micromagnetic simulations, we interpret the effect of pressure in terms of changes to the relative strengths of terms appearing in an effective Hamiltonian.

\section{Experimental and Computational Details}\label{sec:expdet}
Polycrystalline samples of \gavs\ were synthesized and characterized as described in Refs.~\cite{franke2018magnetic,stefancic2020establishing}.
Measurements of the magnetization were made using a Quantum Design Magnetic Property Measurement System.
The sample was loaded into a EasyLab Mcell pressure cell which allowed pressures of up to 1~GPa.
The pressure was measured \textit{in-situ} by monitoring the superconducting transition of a tin manometer. Temperature scan measurements were performed in an applied magnetic field (either 10~mT or 5~T) on cooling from room temperature.
Field scan measurements were performed on decreasing field at 2~K.

\musr\ measurements~\cite{blundell1999spin,blundell2021muon,sm} of polycrystalline \gavs\ were carried out at the $\mu$E1 beamline of the Swiss Muon Source (S$\mu$S), Paul Scherrer Institute, Switzerland, using the GPD instrument.
Both zero-field (ZF) measurements (where no external magnetic field is applied), and transverse-field (TF) measurements (where an external magnetic field is applied perpendicular to the initial muon-spin polarization) were performed.
Polycrystalline samples were loaded in a double-wall piston cylinder cell made of MP35N material~\cite{khasanov2016high,shermadini2017low}, using Daphne oil (7373) as a pressure-transmitting medium, which was mounted in a Janis cryostat.
Data analysis was carried out using the WiMDA program~\cite{pratt2000wimda} and made use of the MINUIT algorithm~\cite{james1975minuit} via the iminuit~\cite{iminuit} Python interface for global refinement of parameters.
We have used the muon stopping sites calculated in Ref.~\cite{franke2018magnetic}, along with the MuESR code~\cite{bonfa2018introduction}, to perform simulations of magnetic field distributions as seen with \musr\ in \gavs.

Micromagnetic simulations were carried out using the ubermag package~\cite{beg2017user,ubermag}.
A $512\times512\times1$ grid of cells (side length 0.8~nm) with periodic boundary conditions in all directions was simulated.
The $C_{nv}$ crystal class was used, with micromagnetic parameters based on previous experimental and computational work~\cite{kezsmarki2015neel,ehlers2016exchange,padmanabhan2019optically}.
We set exchange $\mathcal{A}=0.05975$~pJ/m, DMI $D_0=0.03057$~mJ/m$^2$, magnetocrystalline anisotropy $K_0=16$~kJ/m$^3$, and saturation magnetization $M_\text{s}=33.07$~kA/m.
The easy axis $\vect{u}$ was out of the plane of the simulation, aligned with the magnetic field direction.

\section{Results \& Discussion}
\begin{figure}
	\centering
	\includegraphics[width=\linewidth]{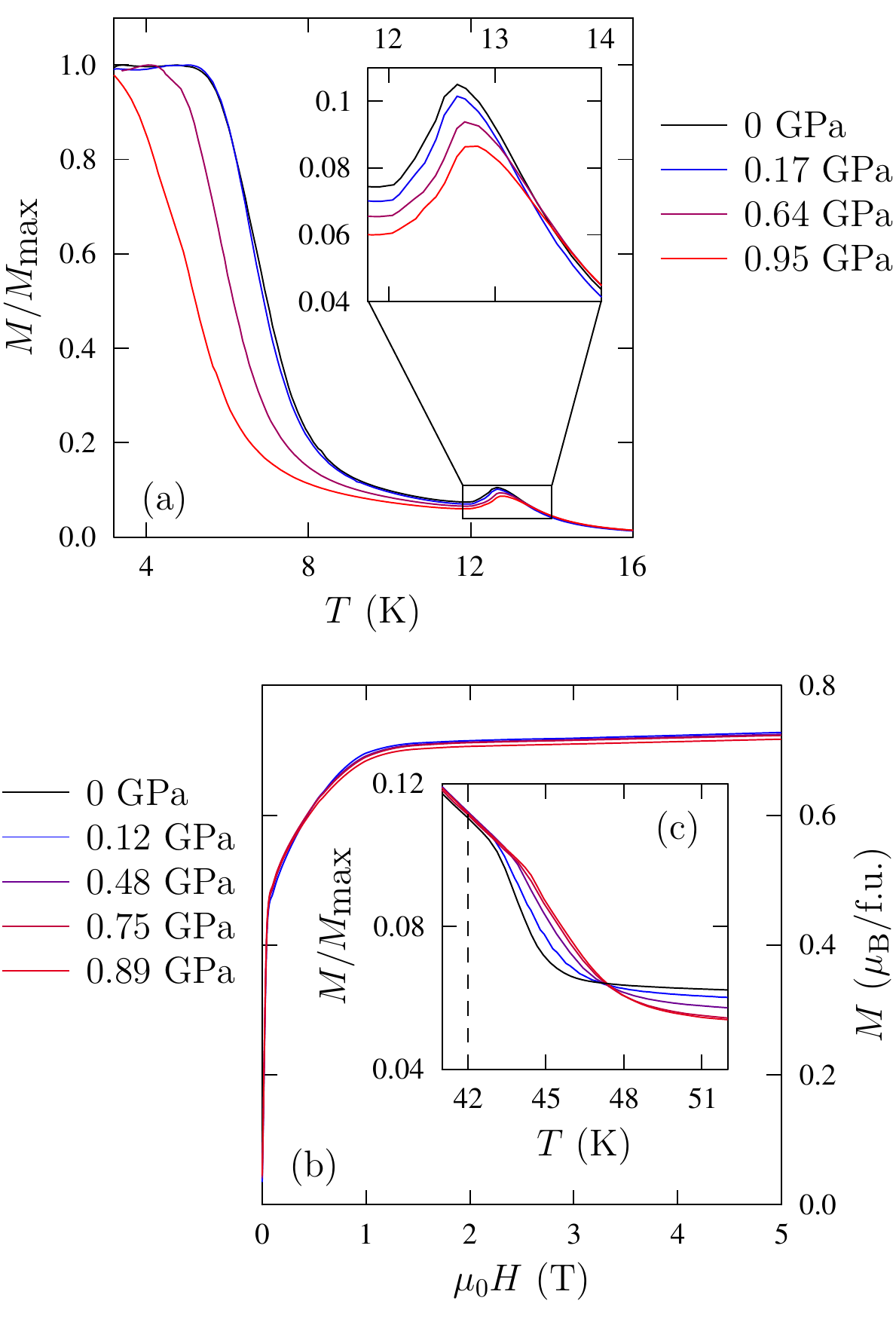}
	\caption{Magnetization measurements of \gavs\ under the application of various external pressures. (a) A temperature scan in an applied field of 10~mT. Data are normalized to the maximum magnetization at each pressure. Inset shows an enlargement of the region around $T_\text{c}~\simeq$~13~K. (b) A field scan at 2~K. (c) The change in magnetization associated with the Jahn-Teller induced structural phase transition, measured in an applied field of 5~T. The 0~GPa measurements are performed without the pressure cell. As in (a), the data are normalized. The dashed line marks $T_\text{JT}$ at $P=0$~GPa according to Ref.~\cite{holt2020structure}.}
	\label{fig:magnetisation}
\end{figure}

The results of magnetization measurements made at several applied pressures are shown in Fig.~\ref{fig:magnetisation}.
We start by considering the measurements performed under the application of a small $\mu_0H=10$~mT external field [Fig.~\ref{fig:magnetisation}(a)] with the intention of studying the system close to the magnetic ground state.
In these measurements, the pressure cell adds a significant background to the measured magnetization.
The data in Fig.~\ref{fig:magnetisation}(a) are therefore normalized to the maximum magnetization $M_{\mathrm{max}}$ at each pressure.
There is no significant change in this quantity with pressure, which appears to vary randomly by a few percent.
As the temperature is decreased, the measurements share the same features: (i) a small peak at $T~\simeq$~13~K, (ii) a range of $T$ where $M/M_{\mathrm{max}}\lesssim 0.2$, and (iii) a rapid increase in the magnetization at low $T$.

We  attribute each of the three features to different physical origins.
(i) The peak at $T~\simeq$~13~K marks $T_\text{c}$, which increases very slightly with pressure, although this change is small compared to other features.
As $T_\text{c}$ is set, predominantly, by the strength of the exchange interaction $\mathcal{A}$, we infer that pressure does not significantly change this parameter.
(ii) The range of $T$ where $M/M_{\mathrm{max}}$ is small corresponds to the C state, where the rotating spin structure leads to almost zero magnetization.
(iii) The increase in magnetization as $T$ decreases further is consistent with the crossover to the FM* phase. The sample magnetization does not increase further once the FM* state is stabilized over the entire sample.
Pressure has a marked effect on the temperature at which the magnetization increase occurs, suggesting that as the pressure is increased, the C to FM* crossover occurs at successively lower $T$.
There are two changes to the spin Hamiltonian that might explain this behavior: an increase in the strength of the DMI (making the twisting C state more preferable compared to the FM* state), or a decrease in the easy-axis anisotropy (making the spins less likely to align along a particular direction).
These two possibilities are discussed in more detail below.

Measurements were also performed at $T=2$~K as a function of applied field [Fig.~\ref{fig:magnetisation}(b)].
These results are not normalized, and show the magnetization saturates at effectively the same value regardless of applied pressure.
There is a slight reduction in magnetization with increasing pressure, however this is very small ($<1.7\%/$GPa), and is therefore hard to unambiguously say whether this is an artifact from the pressure cell.
The pressure cell does have a significant impact on the magnetization, hence the absolute values extracted should be treated with caution; the measured magnetization is significantly suppressed compared to measurements of the same samples measured outside of the pressure cell (where saturation is about $0.85~\mu_\text{B}/$f.u.).

Figure~\ref{fig:magnetisation}(c) shows the temperature dependence of the magnetization under the application of a $\mu_0H=5$~T external field, where the system is in a field-polarized state below $T_\text{c}$, around the temperature below which the Jahn-Teller distortion occurs, $T_\text{JT}$($=42$~K at $P=0$~\cite{holt2020structure}).
As in Fig.~\ref{fig:magnetisation}(a), the data have been normalized.
The effect of the pressure cell (proportionally larger above $T_\text{c}$ due to the small sample signal) results in a pressure-independent shift of features to a slightly higher $T$.
There appears to be a systematic increase in $T_\text{JT}$ as $P$ increases (around 1.3~K/GPa at these pressures).
Reference~\cite{wang2021semiconducting} reports that above 35~GPa, at room temperature, the system adopts the $Imm2$ structure.

To further understand the crystal structures, and the effect of pressure, we can consider related chemically-substituted systems, where the effect of substitution is similar to that of an external pressure.
In GaMo$_4$Se$_8$, when the Jahn-Teller distortion is decomposed into different normal modes, it has been shown that~\cite{routledge2021mode} different amplitudes of these modes results in either the $F\bar{4}3m$ phase (stabilized in \gavs\ at low $P$, low $T$), or $Imm2$ phase (stabilized in \gavs\ at high $P$, high $T$).
Further, Ref.~\cite{schueller2020structural} reports that GaMo$_4$Se$_8$ exhibits coexistence of the ground state $F\bar{4}3m$ structure and the metastable $Imm2$ structure at low $T$.
In contrast, in GeV$_4$S$_8$, the structure below $T_\text{JT}$ is $Imm2$~\cite{muller2006magnetic,bichler2008structural}.

One possible explanation for the changes in the structure of \gavs\ with pressure is that the transition observed at room temperature, $P\simeq35$~GPa, is not the same as that observed at $T_\text{JT}$ in the absence of applied pressure.
An alternative, perhaps simpler, explanation could be that (i) $T_\text{JT}$ increases with pressure in \gavs\ such that $T_\text{JT}$ is above room temperature for pressures above 35~GPa, and that (ii) changes to the nature of the distortion results in the realization of the $Imm2$ phase.
For this scenario there would either need to be an increase in the rate of change of $T_\text{JT}$ with pressure (extrapolating the rate from our data is not sufficient), or a discontinuous change in $T_\text{JT}$ due to some significant change in the behavior of the system (which could be related to the realization of the $Imm2$ phase).

\begin{figure}
	\centering
	\includegraphics[width=\linewidth]{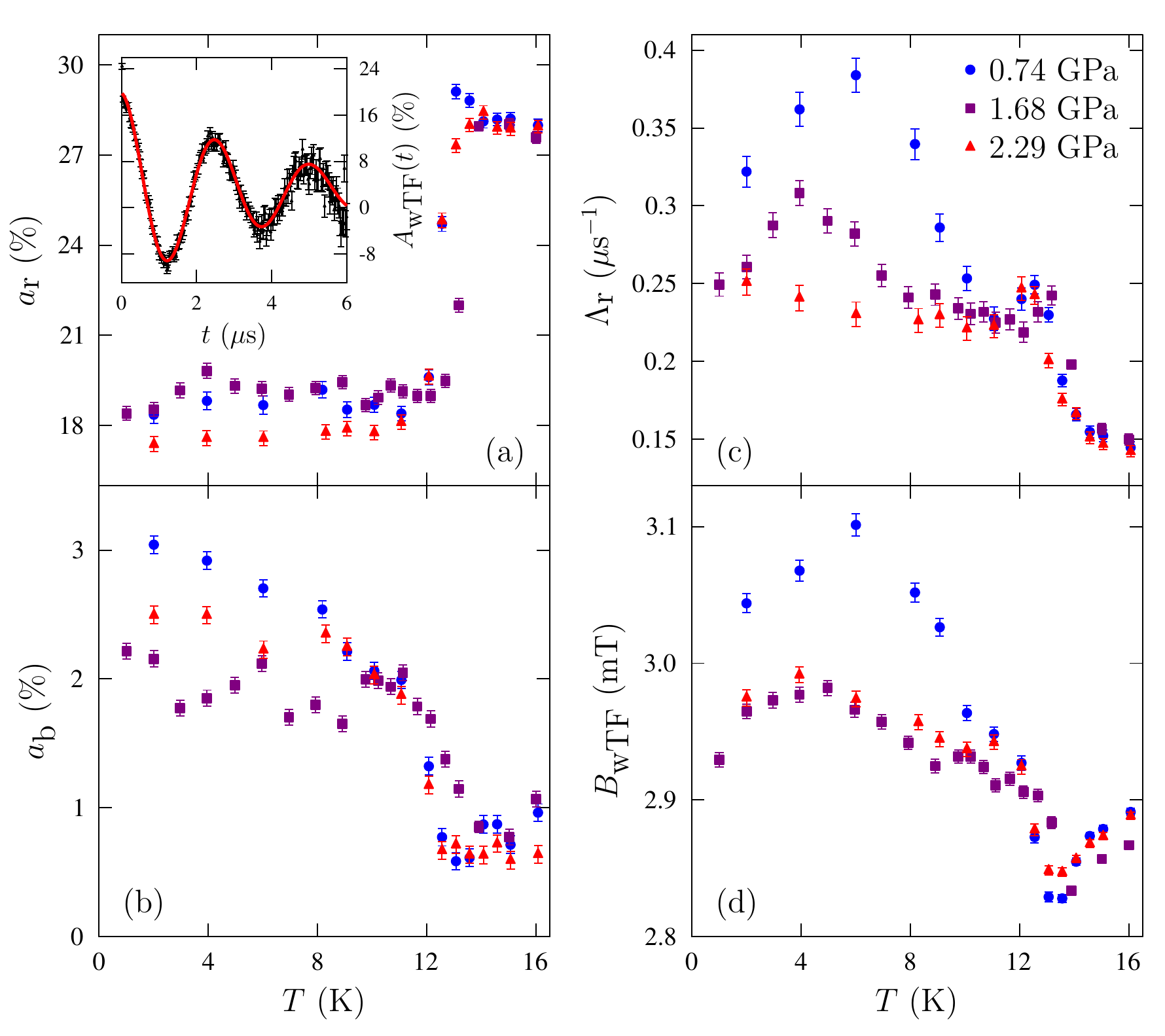}
	\caption{Parameters extracted through fitting wTF \musr\ measurements of \gavs\ under various indicated external pressures in a applied field of $3$~mT. Asymmetries are shown on the left (a--b), with $\Lambda_\text{r}$ and $B_\text{wTF}$ on the right (c--d) describing the effects of residual fields on muons outside the sample, or in positions of near cancellation of the internal field. The inset of (a) shows representative data measured at 2.29~GPa, 0.4~K, with the corresponding fit.}
	\label{fig:musrwtf}
\end{figure}

To further probe the magnetic phase diagram of \gavs, we performed \musr\ measurements with the application of a weak transverse field (wTF), and in zero applied magnetic field (ZF).
Previous \musr\ measurements in this regime~\cite{franke2018magnetic,hicken2020magnetism} revealed a complicated state of affairs with two classes of muon sites sensitive to both static and dynamic effects, with behavior likely resulting from the evolution of the magnetic structure and the magnetic domains with temperature. 
We first consider the wTF measurements, useful in determining the nature of magnetic transitions, which were performed in 3~mT applied perpendicular to the initial muon-spin direction.
The measured asymmetry spectra $A_\text{wTF}\left(t\right)$ (example spectra are shown in the inset of Fig.~\ref{fig:musrwtf}(a), and in the Supplemental Material~\cite{sm}) have two clear contributions from the precession of the muon-spin, one from the response of muons predominantly sensitive to the external wTF (i.e.\ stopped outside the sample, or in positions where the internal field is small), and the other from muons that stop in sites dominated by the internal field in the material.
The contribution from these second class of muon sites track the results we obtain with ZF \musr\ discussed later, but with an expected larger uncertainty than the ZF measurement.
We parameterize the data with
\begin{equation}
	A_\text{wTF}\left(t\right)=a_r\cos\left(\gamma_\mu B_\text{wTF}\right)\exp\left(-\Lambda_\text{r} t\right) + a_\text{b},
\end{equation}
for all $T$, excluding the first 0.2~$\mu$s of data from the fits (where the spectra is dominated by the rapid precession and relaxation from muons stopping in the sites sensitive to the internal field).
This captures contributions from all muon sites for $T>T_{\mathrm{c}}$, but the oscillating component loses the contribution from those muon sites experiencing the large, internal magnetic field resulting from magnetic order below $T_{\mathrm{c}}$. 
The results from this parameterization can be seen in Fig.~\ref{fig:musrwtf}.

The relaxing asymmetry, $a_\text{r}$ corresponds to muons stopping at sites where the field is $B_\text{wTF}$.
The associated relaxation rate $\Lambda_\text{r}$ can arise due to both static disorder (leading to a distribution of local magnetic fields at the muon sites, whose width determines the relaxation rate) and dynamic fluctuations on the muon timescale (where the amplitude and rate of the fluctuating field determines the relaxation rate).
Note that the observed exponential relaxation usually corresponds to dynamic fluctuations.
We find that $a_\text{r}$ decreases below $T_\text{c}$ as expected for a magnetic transition [Fig.~\ref{fig:musrwtf}(a)], where we usually see the loss of the signal from the magnetically ordered component of the sample.
In this regime the temperature-dependent $a_\text{b}$ [Fig.~\ref{fig:musrwtf}(b)] mainly captures the contribution from muons with their spins directed along the large, static internal field in the ordered regime (and hence do not precess).
The temperature evolution of this component is slightly unusual, but likely reflects dynamic fluctuations that freeze out as $T\rightarrow0$, leading to a recovery of $a_\text{b}$.
The effect of pressure does not change $a_\text{r}$ within the statistical accuracy of the data and we see no evidence for a pressure-induced magnetic phase separation from these results.
Despite the gradual temperature evolution of $a_\text{b}$, these results, taken with the previous \musr, suggest the entire sample shows long-range magnetic order below $T_\text{c}$ at all pressures.

We now consider $\Lambda_\text{r}$ and $B_\text{wTF}$ [Figs.~\ref{fig:musrwtf}(c--d)], whose behavior below $T_{\mathrm{c}}$ reflects muons stopping outside the sample or in positions where the internal field is small. 
Considering first the low-temperature FM$^*$ to C transition, at 0.74~GPa, both $\Lambda_\text{r}$ and $B_\text{wTF}$ show a local maximum coincident with the crossover, consistent with magnetometry results and a similar peak in the relaxation rate observed at ambient pressure~\cite{hicken2020magnetism}.
As $P$ is increased, these peaks are pushed to lower $T$ (consistent with the magnetization results), before the peak becomes impossible to resolve at 2.29~GPa, suggesting the C state persists down to lower $T$ than measured.
A peak in $\Lambda_\text{r}$ is not unexpected for a phase transition or crossover, however, the peak in $B_\text{wTF}$ is more unusual.
We suggest that since, as $T$ increases, the anisotropy in \gavs\ decreases~\cite{ehlers2016exchange}, this will lead to a changing orientational preference of the magnetic domains in the sample, altering the magnetic field experienced by the muons outside the samples due to the macroscopic fields from the different grains.
The field experienced by the muons will then peak in the FM$^*$ state when domains are most aligned, which should occur at the highest temperature where the FM$^{*}$ state is realized.
The peak in $B_\text{wTF}$, therefore, coincides with the transition between the FM$^*$ and C states.

\begin{figure}
	\centering
	\includegraphics[width=\linewidth]{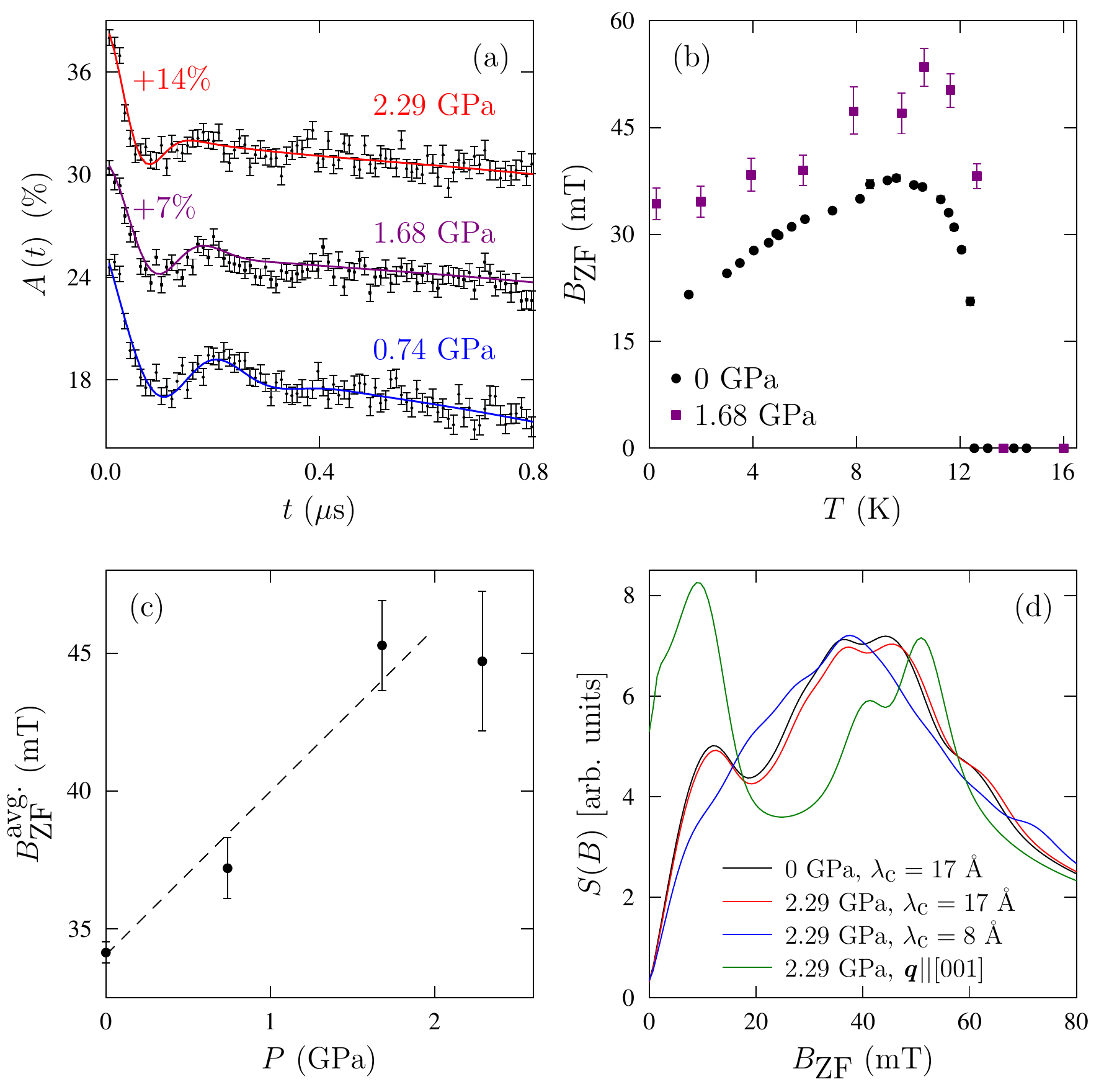}
	\caption{(a) Example ZF \musr\ asymmetry spectra of \gavs\ at various pressures, all at 2~K. The internal field extracted from fitting the 1.68~GPa data is shown in (b), and compared to the zero-pressure result reported in Ref.~\cite{hicken2020magnetism}. All other measured pressures show the same shape~\cite{sm}. The average internal field for $8$~K~$<T<T_\text{c}$ (i.e. in the C phase at all pressures) at each pressure is shown in (c). (d) Simulations of the internal field distribution as seen with ZF \musr\ for the cycloidal state in \gavs\ under various different possible conditions.}
	\label{fig:musrzf}
\end{figure}

ZF \musr\ measurements were also performed as a function of pressure, with example spectra shown in Fig.~\ref{fig:musrzf}(a).
Below $T_\text{c}$ the asymmetry spectra $A\left(t\right)$ are well described by
\begin{multline}
	A\left(t\right)=a_1\cos\left(\gamma_\mu B_\text{ZF}+\phi\right)\exp\left(-\sigma_1^2t^2\right) \\+ a_2\exp\left(-\Lambda_2 t\right) + a_3\exp\left(-\sigma_3^2t^2\right) + a_4 .
\end{multline}
Above $T_\text{c}$ only the $a_3$ and $a_4$ terms are needed.
The first term accounts for muons that stop in sites with a component of $B_\text{ZF}$ perpendicular to the initial muon-spin, leading to coherent precession, whereas the second term accounts for muons that stop with the initial muon-spin parallel to the local field and relax due to dynamics in the local field distribution.
In these measurements, there is a significant contribution from the pressure cell, which is captured by the third and fourth terms, leading to several parameters that are temperature-independent: $a_1/(a_1+a_2+a_3)$, $a_2/(a_1+a_2+a_3)$, $a_3/(a_1+a_2+a_3)$.
We find that $a_1+a_2+a_3$ increases with $T$, and $\sigma_3$, and $a_4$ predominantly capture the temperature-dependence of the cell.
We are unable to extract any useful information from $\Lambda_2$ (which is determined by dynamic fluctuations).
We also find that $\sigma_1$ (likely to be determined by a mixture of static disorder and residual dynamics) is temperature-independent, and hence was refined simultaneously at all $T$.
The most interesting parameter is therefore the internal field $B_\text{ZF}$ [Fig.~\ref{fig:musrzf}(b)].
(The behavior we observe here contrasts with our previous work~\cite{hicken2020magnetism} where two internal fields were detected, one higher field, broadly matching the low-pressure $B_\text{ZF}$ found here, and one lower field, which is not observable in this configuration, likely due to the competing signal from the pressure cell.)
At all pressures, $B_\text{ZF}$ increases slightly with increasing $T$, as previously observed~\cite{hicken2020magnetism} (shown for comparison), before collapsing to zero above $T_\text{c}$ as expected.

\begin{table}
	\caption{Method of extracting transition temperatures shown in Fig.~\ref{fig:phasediagram} from different measurement types.}
	\centering
	\begin{tabular}{c|c|c|c}
		\hline
		\hline
		Measurement & Parameter & Transition & Extracted feature \\
		\hline
		\hline
		Magnetization & $M$ & $T_\text{c}$ & Local maximum \\
		& $M$ & FM$^*\rightarrow$~C & $M=M_\text{max}/2$ \\
		\hline
		wTF \musr & $\Lambda_\text{r}$ & $T_\text{c}$ & Local maximum \\
		& $\Lambda_\text{r}$ & FM$^*\rightarrow$~C & Local maximum \\
		& $B_\text{wTF}$ & $T_\text{c}$ & Local minimum \\
		& $B_\text{wTF}$ & FM$^*\rightarrow$~C & Local maximum \\
		\hline
		ZF \musr & $B_\text{ZF}$ & $T_\text{c}$ & $B_\text{ZF}=0$ \\
		\hline
		\hline
	\end{tabular}
	\label{tab:features}
\end{table}

Despite no notable changes in the saturation magnetization with pressure [Fig.~\ref{fig:magnetisation}(b)], the local internal field significantly increases as the pressure is increased, as shown in Fig.~\ref{fig:musrzf}(c).
This suggests a change to the structure of the C state that does not lead to a change in the net magnetization of the FM$^*$ state.
To explore possibilities that explain this, we consider the changes to the Hamiltonian previously suggested: an increase in the DMI, or a decrease in the anisotropy.
An increase in DMI with pressure would be expected to lead to a decrease in the cycloidal period (as the wavelength $\lambda_\text{C}\propto \mathcal{A}/D$).
We can simulate the effect this change in the period would have on the distribution of fields seen at the muon site following the approach in Ref.~\cite{hicken2020magnetism}, which was shown to describe the ZF \musr\ data.
The values of $\sigma_1$ and $\Lambda_2$ are the same order of magnitude as the comparable parameters $\Lambda_i$ in Ref.~\cite{hicken2020magnetism}, suggesting the width of the distribution of magnetic fields at the muon site remains similar upon application of pressure.
We therefore conclude that the simulations are likely still good descriptors of the ZF \musr\ data.

The simulated distribution~\cite{hicken2020magnetism} produced for the experimentally observed C state ($\lambda_\text{C}=17$~nm, and the $\vect{q}$-vector perpendicular to the [001] direction~\cite{ruff2015multiferroicity}) is shown in Fig.~\ref{fig:musrzf}(d).
We first consider the effect on the distribution that the small change in lattice parameters arising due to the pressure would have.
Taking the change in cell volume from Ref.~\cite{wang2021semiconducting}, and assuming the muon sites remain unchanged (a reasonable assumption given the minimal change to the lattice parameters), we can see that the spectrum is almost unchanged.
Reducing the cycloidal period to $\lambda_\text{C}=8$~nm (equivalent to approximately doubling the  relative strength of the DMI) changes the form, but not the position of the features (i.e. $B_\text{ZF}$) in the spectrum.
It is therefore unlikely that this can explain the observed increase in internal field.
Further, as changes to $\mathcal{A}$ would also result in changes to $\lambda_\text{C}$, these results also support the conclusion from the magnetization measurements that $\mathcal{A}$ does not significantly change with pressure.

One possible effect of a reduction in the anisotropy as the pressure is increased could be a changed preference for the direction of the $\vect{q}$-vector of the cycloidal state.
In Fig.~\ref{fig:musrzf}(d) we show that, if the $\vect{q}$-vector were to point along the [001] direction, the larger-field peak is shifted to higher fields, as seen experimentally.
We note that changing the $\vect{q}$-vector was the only way we could find to simulate the observed increase in internal field without changing the net magnetization.
Whilst we cannot say for certain that this is the only orientation of the $\vect{q}$-vector that would reproduce this result, we found no other feasible direction when randomly sampling, or testing specific high symmetry directions.
We, therefore, suggest that a change in the anisotropy is a possible explanation for both the reduction in temperature of the C to FM* crossover, and the increase in internal field with increasing pressure.

\begin{figure}
	\centering
	\includegraphics[width=0.7\linewidth]{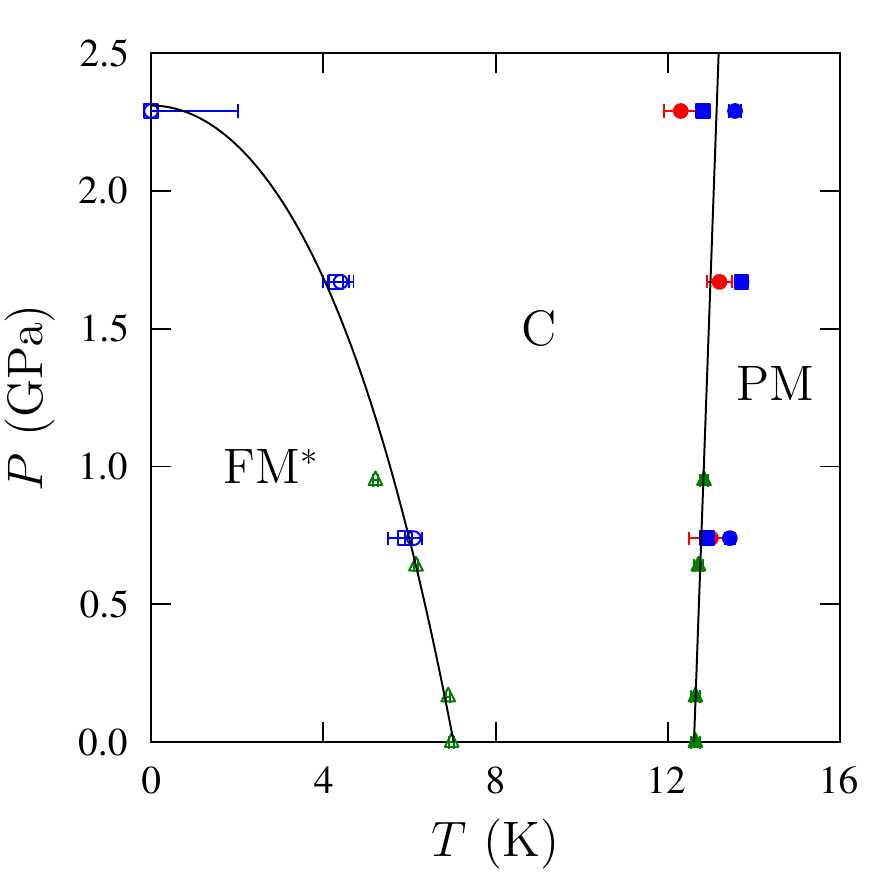}
	\caption{The $P$--$T$ dependence of \gavs\ based on magnetization (green) and \musr\ (ZF: red; wTF: blue) measurements. Closed symbols indicate $T_\text{c}$, above which the system is paramagnetic (PM), and open symbols represent the ferromagnetic-like (FM$^*$) to cycloidal (C) transition. Circles indicate the point comes from the extracted value of the internal field [Fig.~\ref{fig:musrwtf}(d), Fig.~\ref{fig:musrzf}(b)], squares represent points from a feature in $\Lambda_\text{r}$ [Figs.~\ref{fig:musrwtf}(c)], and triangles represent points from magnetization [Fig.~\ref{fig:magnetisation}(a)]. The points are extracted as detailed in Table~\ref{tab:features}.}
	\label{fig:phasediagram}
\end{figure}

We have summarized our magnetization and \musr\ results with a suggested $P$--$T$ phase diagram in the absence of an applied field, Fig.~\ref{fig:phasediagram}.
We have extracted the transition temperatures from the different types of measurement as shown in Table~\ref{tab:features}.
We find that $T_\text{c}$ slightly increases with applied pressure, whereas the C to FM* crossover decreases in $T$ as the pressure increases.
The  error bar on the highest pressure, low temperature point represents the minimum temperature we were able to measure.

\begin{figure*}
	\centering
	\includegraphics[width=\linewidth]{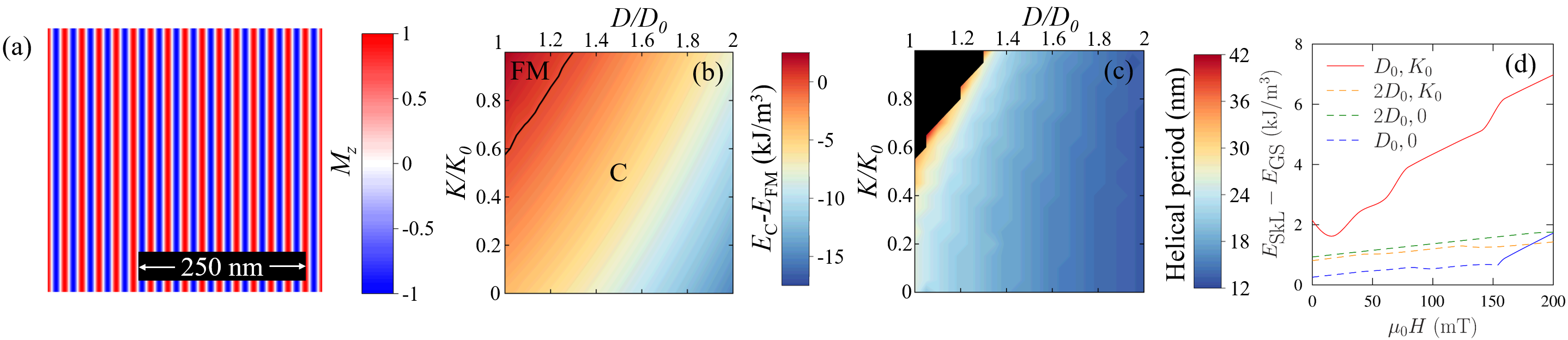}
	\caption{Results of micromagnetic simulations using the parameters in Sec.~\ref{sec:expdet}. (a) An example cycloidal (C) state. (b) Difference in energies between the C and ferromagnetic (FM) states as a function of $D$ and $K$, with the preferred cycloidal period shown in (c). (d) Energies of the skyrmion lattice (SkL) state compared to the ground state (either FM, solid line, or C, dashed line) as a function of applied magnetic field for different combinations of $D$ and $K$.}
	\label{fig:micromagnetics}
\end{figure*}

To further study the effects on the magnetic states of these possible changes in the DMI and/or anisotropy, we have performed micromagnetic simulations.
We use the Hamiltonian, 
\begin{multline}
	\hham = -\mathcal{A}\vect{m}\cdot\nabla^2\vect{m} + D\left(\vect{m}\cdot\nabla m_z-m_z\nabla\cdot\vect{m}\right) \\- K\left(\vect{m}\cdot\vect{u}\right)^2 - \mu_0M_\text{s}\vect{m}\cdot\vect{H} ,
\end{multline}
which is appropriate for a $C_{nv}$ crystal system, with the micromagnetic parameters defined in Sec.~\ref{sec:expdet}.
The exchange $\mathcal{A}$ term attempts to locally align $\vect{m}$ whereas the DMI $D$ term prefers periodic rotation of $\vect{m}$.
The anisotropy $K$ term gives a preference for alignment along the easy axis $\vect{u}$, and the final term lowers the energy when $\vect{m}$ is aligned along the external field $\vect{H}$.
We initialize different magnetic configurations before allowing the micromagnetic solver to find a local energy minimum, allowing us to probe the energy of many possible states.

In the absence of an applied field we find that the FM state is the most energetically favorable, with the C state only slightly higher in energy.
(Note that, on the length scale of this simulation, each micromagnetic cell contains around four V$_{4}$ clusters, hence the FM$^*$ and FM state are indistinguishable.)
An example of the C state simulated can be seen in Fig.~\ref{fig:micromagnetics}(a).
We study the effect of changing the $D$ and $K$ parameters, keeping the other parameters fixed, and identify the difference in energy between the lowest energy C state (i.e.\ that which has the preferred C period) and the FM state.
This is shown in Fig.~\ref{fig:micromagnetics}(b), where the lowest energy state is marked.
As expected, both increasing $D$ and decreasing $K$ change the preferred state from FM to C.
The minimum energy cycloidal period is shown in Fig.~\ref{fig:micromagnetics}(c), and shows that changes in $\lambda_\text{C}$ are mainly correlated with changes in $D$.

Having established micromagnetics can reproduce the observed states, we can examine the effect of applying a magnetic field.
We consider the energy of the skyrmion lattice (SkL) state compared to the ground state (either FM or C, depending on the micromagnetic parameters chosen).
We find that, for $D_0$, $K_0$, (i.e. the zero-temperature parameters we expect in the absence of any applied pressure) the SkL state is hard to stabilize, with very few initialized SkL states retaining this character.
The SkL state is never found to be lower in energy than the C or FM state, as expected from the measured phase diagram, where the SkL is only stabilized via thermal fluctuations above $T=0$ in an applied field, typically over a range of 2--3~K below $T_\text{c}$.
Changing the values of $D$ and $K$ makes stabilizing the SkL much more probable, with many more initialized SkL states staying in that configuration, and the energy of the skyrmion state above the ground state (normally C, apart from at high field in the $D=D_0$, $K=0$ case where the ground state becomes FM) is much reduced.
These micromagnetic simulations therefore suggest that, if the effect of changing pressure is a change in the $D$ or $K$ parameters, the SkL should still be stabilized at elevated temperatures, as it is in the absence of an applied pressure, but with it likely to persist down to lower $T$.
However, since the SkL is still never found to be the lowest energy state it is unlikely to ever be stabilized at zero temperature.

\begin{figure}
	\centering
	\includegraphics[width=\linewidth]{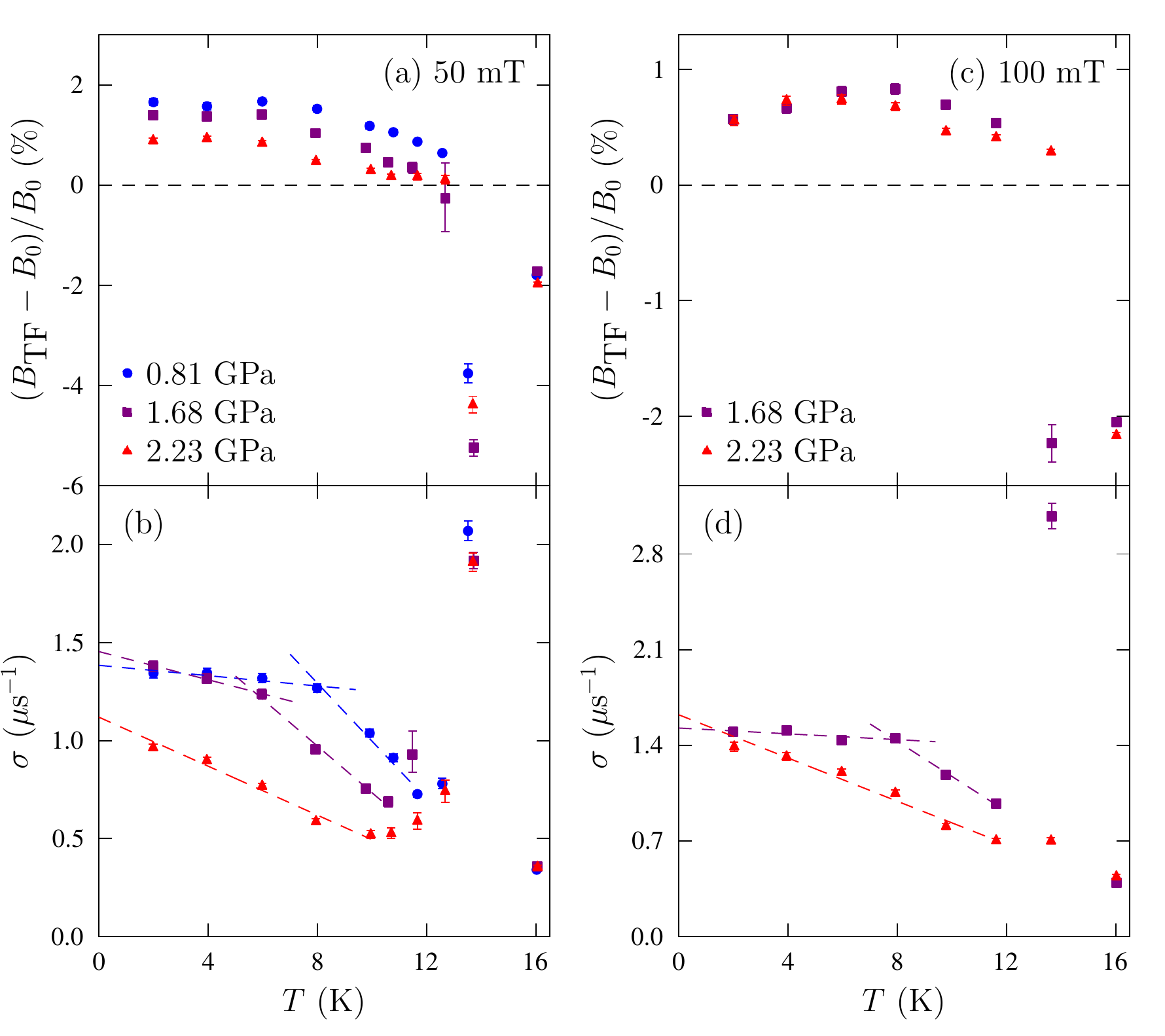}
	\caption{Parameters extracted from fitting of TF \musr\ measurements of \gavs\ at different applied pressures and magnetic fields. Dashed lines in (b) and (d) are guides to the eye.}
	\label{fig:musrhighfield}
\end{figure}

TF \musr\ has previously been shown to be able to identify the SkL phase in \gavs\ via an increase in the internal field and the associated relaxation rate~\cite{hicken2020magnetism}.
We have performed TF \musr\ measurements of \gavs\ at 50~mT and 100~mT (fields which stabilize a SkL in the absence of applied pressure) under the application of pressure.
The asymmetry spectra $A_\text{TF}\left(t\right)$ are well described by
\begin{equation}
	A_\text{TF}\left(t\right)=\sum_{i=1}^2a_i\cos\left(\gamma_\mu B_{\text{TF},i}+\phi_i\right)\exp\left(-\sigma_i^2t^2\right) + a_\text{b} .
\end{equation}
The $i=2$ term captures the effects of muons stopping in the pressure cell when the spins subsequently precess, hence we employed simultaneous refinement of the associated temperature-independent parameters: $B_{\text{TF},2}$, $\phi_2$, and $\sigma_2$.
The total asymmetry $a_1+a_2$ increases at $T_\text{c}$ as expected, with $a_\text{b}$ capturing temperature dependence that appears to be coming from the cell (presumably from muons that stop at sites in the cell where the total magnetic field is zero).
The first term, sensitive to the magnetism in \gavs, has a temperature-independent $\phi_1$ which was simultaneously refined at all temperatures.
The remaining parameters $B_{\text{TF},1}$ and $\sigma_1$ are shown in Fig.~\ref{fig:musrhighfield}.

At all applied pressures, the behavior is broadly similar in both parameters.
We first consider the measured internal field, Figs.~\ref{fig:musrhighfield}(a,c).
At 50~mT, there is an enhancement of $B_\text{TF}$ compared to the applied field, as previously seen in the absence of applied pressure~\cite{hicken2020magnetism}.
There is no large increase in this parameter over a limited temperature range that would signify the stabilization of the SkL in that region, as has previously been observed at ambient pressure.
Instead, the internal field seems enhanced for all $T<T_\text{c}$.
Comparatively, at 100~mT the enhancement from the applied field is smaller.

Considering now the relaxation rate $\sigma$ [Figs.~\ref{fig:musrhighfield}(b,d)], we resolve two clear regimes of behavior  at 0.81~GPa and 1.68~GPa in both the 50~mT and 100~mT measurements, as marked by dashed lines.
The crossover $T$ between these two regimes decreases as the pressure is increased, following a similar trend seen earlier for the FM$^*$ to C transition.
By 2.23~GPa it is difficult to split the behavior into separate regimes, again as seen in the zero-field data.
We conclude, therefore, that pressure acts to reduce the characteristic temperature scales of the magnetic field fluctuations in this part of the phase diagram.
We stress that we are unable to resolve those features seen at ambient pressure~\cite{hicken2020magnetism} that unambiguously mark the onset of the SkL.
However, since fluctuations in the field distribution experienced by the muon have previously been shown to track the onset of the skyrmion phase, we might speculate that these data are consistent with the SkL phase extending to lower $T$, as anticipated from the micromagnetic results.
The phase diagram would therefore look more like that of GaV$_4$Se$_8$, which has a SkL stable down to low $T$~\cite{fujima2017thermodynamically}.

\section{Conclusion}
Our magnetization and \musr\ measurements of \gavs\ reveal that the application of pressure lowers the crossover temperature of the C to FM* transition.
By 2.29~GPa, we cannot observe this crossover, and suggest that \gavs\ remains in the C state down to the lowest measured temperatures.
On increasing the applied pressure, the internal field as measured by \musr\ increases without any corresponding increase in the magnetization.
Through simulations of the distribution of magnetic fields at the muon stopping sites, we suggest this occurs due to a reorientation of the direction of the $\vect{q}$-vectors due to a reduction in the magnetic anisotropy of the system.
Investigation of this phenomena with a $\vect{q}$-resolved experimental probe would be highly beneficial in a future study.
Micromagnetic simulations suggest further that a decrease in the magnetic anisotropy should lead to an increase in stability of the SkL state in \gavs.

There are multiple reasons that application of pressure may change the anisotropy of the system.
The effect of pressure might be to make the underlying electronic structure more isotropic, hence explaining the above results.
Further, the appearance of an easy-axis in the system is set by the direction of the Jahn-Teller distortion; given our results on the increase of the temperature of the Jahn-Teller distortion with pressure, and the possibility of the changing nature of this distortion, it seems highly likely that the anisotropy would also have corresponding changes with pressure.
Given the possible different scenarios we set out to explain the changes to the Jahn-Teller distortion with pressure, experiments to determine the crystal structure of \gavs\ as a function of both temperature and pressure are an area of research that should be explored.

Our work shows the sensitivity of the magnetism in \gavs\ to applied pressures, and demonstrates that changes in both the magnetic phase diagram and underlying magnetic interaction strengths can be expected.
This provides the opportunity to tailor the magnetic interactions of this system, and perhaps related systems in the GaV$_4$S$_{8-y}$Se$_y$ series, and investigate the effects on the magnetism.
Understanding the changes these parameters make is of great importance for understanding the behavior of these complex magnetic states, which is essential if they are to be used for  applications.

\section*{Acknowledgments}
Part of this work was carried out at the Swiss Muon Source, Paul Scherrer Institute, Switzerland; we are grateful for the provision of beamtime.
We are grateful for computational support from Durham Hamilton HPC.
The project was funded by EPSRC (UK) (Grant Nos.: EP/N032128/1 and EP/N024028/1).
M. N. Wilson acknowledges the support of the Natural Sciences and Engineering Research Council of Canada (NSERC).
Research data from this paper will be made available via Durham Collections at \textcolor{red}{XXX}.

\bibliography{bib}

\begin{thebibliography}{32}%
\makeatletter
\providecommand \@ifxundefined [1]{%
 \@ifx{#1\undefined}
}%
\providecommand \@ifnum [1]{%
 \ifnum #1\expandafter \@firstoftwo
 \else \expandafter \@secondoftwo
 \fi
}%
\providecommand \@ifx [1]{%
 \ifx #1\expandafter \@firstoftwo
 \else \expandafter \@secondoftwo
 \fi
}%
\providecommand \natexlab [1]{#1}%
\providecommand \enquote  [1]{``#1''}%
\providecommand \bibnamefont  [1]{#1}%
\providecommand \bibfnamefont [1]{#1}%
\providecommand \citenamefont [1]{#1}%
\providecommand \href@noop [0]{\@secondoftwo}%
\providecommand \href [0]{\begingroup \@sanitize@url \@href}%
\providecommand \@href[1]{\@@startlink{#1}\@@href}%
\providecommand \@@href[1]{\endgroup#1\@@endlink}%
\providecommand \@sanitize@url [0]{\catcode `\\12\catcode `\$12\catcode
  `\&12\catcode `\#12\catcode `\^12\catcode `\_12\catcode `\%12\relax}%
\providecommand \@@startlink[1]{}%
\providecommand \@@endlink[0]{}%
\providecommand \url  [0]{\begingroup\@sanitize@url \@url }%
\providecommand \@url [1]{\endgroup\@href {#1}{\urlprefix }}%
\providecommand \urlprefix  [0]{URL }%
\providecommand \Eprint [0]{\href }%
\providecommand \doibase [0]{https://doi.org/}%
\providecommand \selectlanguage [0]{\@gobble}%
\providecommand \bibinfo  [0]{\@secondoftwo}%
\providecommand \bibfield  [0]{\@secondoftwo}%
\providecommand \translation [1]{[#1]}%
\providecommand \BibitemOpen [0]{}%
\providecommand \bibitemStop [0]{}%
\providecommand \bibitemNoStop [0]{.\EOS\space}%
\providecommand \EOS [0]{\spacefactor3000\relax}%
\providecommand \BibitemShut  [1]{\csname bibitem#1\endcsname}%
\let\auto@bib@innerbib\@empty
\bibitem [{\citenamefont {Brasen}\ \emph {et~al.}(1975)\citenamefont {Brasen},
  \citenamefont {Vandenberg}, \citenamefont {Robbins}, \citenamefont {Willens},
  \citenamefont {Reed}, \citenamefont {Sherwood},\ and\ \citenamefont
  {Pinder}}]{brasen1975magnetic}%
  \BibitemOpen
  \bibfield  {author} {\bibinfo {author} {\bibfnamefont {D.}~\bibnamefont
  {Brasen}}, \bibinfo {author} {\bibfnamefont {J.~M.}\ \bibnamefont
  {Vandenberg}}, \bibinfo {author} {\bibfnamefont {M.}~\bibnamefont {Robbins}},
  \bibinfo {author} {\bibfnamefont {R.~H.}\ \bibnamefont {Willens}}, \bibinfo
  {author} {\bibfnamefont {W.~A.}\ \bibnamefont {Reed}}, \bibinfo {author}
  {\bibfnamefont {R.~C.}\ \bibnamefont {Sherwood}},\ and\ \bibinfo {author}
  {\bibfnamefont {X.~J.}\ \bibnamefont {Pinder}},\ }\bibfield  {title}
  {\bibinfo {title} {Magnetic and crystallographic properties of spinels of the
  type {A$_x$B$_2$S$_4$} where {A = Al, Ga}, and {B = Mo, V, Cr}},\ }\href
  {https://doi.org/10.1016/0022-4596(75)90141-3} {\bibfield  {journal}
  {\bibinfo  {journal} {Journal of Solid State Chemistry}\ }\textbf {\bibinfo
  {volume} {13}},\ \bibinfo {pages} {298} (\bibinfo {year} {1975})}\BibitemShut
  {NoStop}%
\bibitem [{\citenamefont {Lancaster}(2019)}]{lancaster2019skyrmions}%
  \BibitemOpen
  \bibfield  {author} {\bibinfo {author} {\bibfnamefont {T.}~\bibnamefont
  {Lancaster}},\ }\bibfield  {title} {\bibinfo {title} {Skyrmions in magnetic
  materials},\ }\href@noop {} {\bibfield  {journal} {\bibinfo  {journal}
  {Contemporary Physics}\ }\textbf {\bibinfo {volume} {60}},\ \bibinfo {pages}
  {246} (\bibinfo {year} {2019})}\BibitemShut {NoStop}%
\bibitem [{\citenamefont {K{\'e}zsm{\'a}rki}\ \emph {et~al.}(2015)\citenamefont
  {K{\'e}zsm{\'a}rki}, \citenamefont {Bord{\'a}cs}, \citenamefont {Milde},
  \citenamefont {Neuber}, \citenamefont {Eng}, \citenamefont {White},
  \citenamefont {R{\o}nnow}, \citenamefont {Dewhurst}, \citenamefont
  {Mochizuki}, \citenamefont {Yanai} \emph {et~al.}}]{kezsmarki2015neel}%
  \BibitemOpen
  \bibfield  {author} {\bibinfo {author} {\bibfnamefont {I.}~\bibnamefont
  {K{\'e}zsm{\'a}rki}}, \bibinfo {author} {\bibfnamefont {S.}~\bibnamefont
  {Bord{\'a}cs}}, \bibinfo {author} {\bibfnamefont {P.}~\bibnamefont {Milde}},
  \bibinfo {author} {\bibfnamefont {E.}~\bibnamefont {Neuber}}, \bibinfo
  {author} {\bibfnamefont {L.~M.}\ \bibnamefont {Eng}}, \bibinfo {author}
  {\bibfnamefont {J.~S.}\ \bibnamefont {White}}, \bibinfo {author}
  {\bibfnamefont {H.~M.}\ \bibnamefont {R{\o}nnow}}, \bibinfo {author}
  {\bibfnamefont {C.~D.}\ \bibnamefont {Dewhurst}}, \bibinfo {author}
  {\bibfnamefont {M.}~\bibnamefont {Mochizuki}}, \bibinfo {author}
  {\bibfnamefont {K.}~\bibnamefont {Yanai}}, \emph {et~al.},\ }\bibfield
  {title} {\bibinfo {title} {N{\'e}el-type skyrmion lattice with confined
  orientation in the polar magnetic semiconductor {GaV$_4$S$_8$}},\ }\href
  {https://doi.org/10.1038/nmat4402} {\bibfield  {journal} {\bibinfo  {journal}
  {Nature Materials}\ }\textbf {\bibinfo {volume} {14}},\ \bibinfo {pages}
  {1116} (\bibinfo {year} {2015})}\BibitemShut {NoStop}%
\bibitem [{\citenamefont {Holt}\ \emph {et~al.}(2020)\citenamefont {Holt},
  \citenamefont {{\v{S}}tefan{\v{c}}i{\v{c}}}, \citenamefont {Ritter},
  \citenamefont {Hall}, \citenamefont {Lees},\ and\ \citenamefont
  {Balakrishnan}}]{holt2020structure}%
  \BibitemOpen
  \bibfield  {author} {\bibinfo {author} {\bibfnamefont {S.~J.~R.}\
  \bibnamefont {Holt}}, \bibinfo {author} {\bibfnamefont {A.}~\bibnamefont
  {{\v{S}}tefan{\v{c}}i{\v{c}}}}, \bibinfo {author} {\bibfnamefont
  {C.}~\bibnamefont {Ritter}}, \bibinfo {author} {\bibfnamefont {A.~E.}\
  \bibnamefont {Hall}}, \bibinfo {author} {\bibfnamefont {M.~R.}\ \bibnamefont
  {Lees}},\ and\ \bibinfo {author} {\bibfnamefont {G.}~\bibnamefont
  {Balakrishnan}},\ }\bibfield  {title} {\bibinfo {title} {Structure and
  magnetism of the skyrmion hosting family {GaV$_4$S$_{8-y}$Se$_y$} with low
  levels of substitutions between {$0\leq y\leq0.5$} and {$7.5\leq y\leq8$}},\
  }\href {https://doi.org/10.1103/PhysRevMaterials.4.114413} {\bibfield
  {journal} {\bibinfo  {journal} {Physical Review Materials}\ }\textbf
  {\bibinfo {volume} {4}},\ \bibinfo {pages} {114413} (\bibinfo {year}
  {2020})}\BibitemShut {NoStop}%
\bibitem [{\citenamefont {Ehlers}\ \emph {et~al.}(2016)\citenamefont {Ehlers},
  \citenamefont {Stasinopoulos}, \citenamefont {K{\'e}zsm{\'a}rki},
  \citenamefont {Feh{\'e}r}, \citenamefont {Tsurkan}, \citenamefont {von
  Nidda}, \citenamefont {Grundler},\ and\ \citenamefont
  {Loidl}}]{ehlers2016exchange}%
  \BibitemOpen
  \bibfield  {author} {\bibinfo {author} {\bibfnamefont {D.}~\bibnamefont
  {Ehlers}}, \bibinfo {author} {\bibfnamefont {I.}~\bibnamefont
  {Stasinopoulos}}, \bibinfo {author} {\bibfnamefont {I.}~\bibnamefont
  {K{\'e}zsm{\'a}rki}}, \bibinfo {author} {\bibfnamefont {T.}~\bibnamefont
  {Feh{\'e}r}}, \bibinfo {author} {\bibfnamefont {V.}~\bibnamefont {Tsurkan}},
  \bibinfo {author} {\bibfnamefont {H.~K.}\ \bibnamefont {von Nidda}}, \bibinfo
  {author} {\bibfnamefont {D.}~\bibnamefont {Grundler}},\ and\ \bibinfo
  {author} {\bibfnamefont {A.}~\bibnamefont {Loidl}},\ }\bibfield  {title}
  {\bibinfo {title} {Exchange anisotropy in the skyrmion host {GaV$_4$S$_8$}},\
  }\href {https://doi.org/10.1088/1361-648X/aa4e7e} {\bibfield  {journal}
  {\bibinfo  {journal} {Journal of Physics: Condensed Matter}\ }\textbf
  {\bibinfo {volume} {29}},\ \bibinfo {pages} {065803} (\bibinfo {year}
  {2016})}\BibitemShut {NoStop}%
\bibitem [{\citenamefont {Holt}\ \emph {et~al.}(2021)\citenamefont {Holt},
  \citenamefont {Ritter}, \citenamefont {Lees},\ and\ \citenamefont
  {Balakrishnan}}]{holt2021investigation}%
  \BibitemOpen
  \bibfield  {author} {\bibinfo {author} {\bibfnamefont {S.~J.~R.}\
  \bibnamefont {Holt}}, \bibinfo {author} {\bibfnamefont {C.}~\bibnamefont
  {Ritter}}, \bibinfo {author} {\bibfnamefont {M.~R.}\ \bibnamefont {Lees}},\
  and\ \bibinfo {author} {\bibfnamefont {G.}~\bibnamefont {Balakrishnan}},\
  }\bibfield  {title} {\bibinfo {title} {Investigation of the magnetic ground
  state of {GaV$_4$S$_8$} using powder neutron diffraction},\ }\href
  {https://doi.org/10.1088/1361-648X/abf9bb} {\bibfield  {journal} {\bibinfo
  {journal} {Journal of Physics: Condensed Matter}\ }\textbf {\bibinfo {volume}
  {33}},\ \bibinfo {pages} {255802} (\bibinfo {year} {2021})}\BibitemShut
  {NoStop}%
\bibitem [{\citenamefont {Hicken}\ \emph {et~al.}(2020)\citenamefont {Hicken},
  \citenamefont {Holt}, \citenamefont {Franke}, \citenamefont {Hawkhead},
  \citenamefont {{\v{S}}tefan{\v{c}}i{\v{c}}}, \citenamefont {Wilson},
  \citenamefont {Gomil{\v{s}}ek}, \citenamefont {Huddart}, \citenamefont
  {Clark}, \citenamefont {Lees} \emph {et~al.}}]{hicken2020magnetism}%
  \BibitemOpen
  \bibfield  {author} {\bibinfo {author} {\bibfnamefont {T.~J.}\ \bibnamefont
  {Hicken}}, \bibinfo {author} {\bibfnamefont {S.~J.~R.}\ \bibnamefont {Holt}},
  \bibinfo {author} {\bibfnamefont {K.~J.~A.}\ \bibnamefont {Franke}}, \bibinfo
  {author} {\bibfnamefont {Z.}~\bibnamefont {Hawkhead}}, \bibinfo {author}
  {\bibfnamefont {A.}~\bibnamefont {{\v{S}}tefan{\v{c}}i{\v{c}}}}, \bibinfo
  {author} {\bibfnamefont {M.~N.}\ \bibnamefont {Wilson}}, \bibinfo {author}
  {\bibfnamefont {M.}~\bibnamefont {Gomil{\v{s}}ek}}, \bibinfo {author}
  {\bibfnamefont {B.~M.}\ \bibnamefont {Huddart}}, \bibinfo {author}
  {\bibfnamefont {S.~J.}\ \bibnamefont {Clark}}, \bibinfo {author}
  {\bibfnamefont {M.~R.}\ \bibnamefont {Lees}}, \emph {et~al.},\ }\bibfield
  {title} {\bibinfo {title} {Magnetism and {N{\'e}el} skyrmion dynamics in
  {GaV$_4$S$_{8-y}$Se$_y$}},\ }\href
  {https://doi.org/10.1103/PhysRevResearch.2.032001} {\bibfield  {journal}
  {\bibinfo  {journal} {Physical Review Research}\ }\textbf {\bibinfo {volume}
  {2}},\ \bibinfo {pages} {032001(R)} (\bibinfo {year} {2020})}\BibitemShut
  {NoStop}%
\bibitem [{\citenamefont {Wang}\ \emph {et~al.}(2021)\citenamefont {Wang},
  \citenamefont {Rahman}, \citenamefont {Sun}, \citenamefont {Knill},
  \citenamefont {Zhang}, \citenamefont {Wang}, \citenamefont {Tsurkan},\ and\
  \citenamefont {K{\'e}zsm{\'a}rki}}]{wang2021semiconducting}%
  \BibitemOpen
  \bibfield  {author} {\bibinfo {author} {\bibfnamefont {Y.}~\bibnamefont
  {Wang}}, \bibinfo {author} {\bibfnamefont {S.}~\bibnamefont {Rahman}},
  \bibinfo {author} {\bibfnamefont {E.}~\bibnamefont {Sun}}, \bibinfo {author}
  {\bibfnamefont {C.}~\bibnamefont {Knill}}, \bibinfo {author} {\bibfnamefont
  {D.}~\bibnamefont {Zhang}}, \bibinfo {author} {\bibfnamefont
  {L.}~\bibnamefont {Wang}}, \bibinfo {author} {\bibfnamefont {V.}~\bibnamefont
  {Tsurkan}},\ and\ \bibinfo {author} {\bibfnamefont {I.}~\bibnamefont
  {K{\'e}zsm{\'a}rki}},\ }\bibfield  {title} {\bibinfo {title} {From
  semiconducting to metallic: {Jahn--Teller}-induced phase transformation in
  skyrmion host {GaV$_4$S$_8$}},\ }\href
  {https://doi.org/10.1021/acs.jpcc.0c10527} {\bibfield  {journal} {\bibinfo
  {journal} {Journal of Physical Chemistry C}\ }\textbf {\bibinfo {volume}
  {125}},\ \bibinfo {pages} {5771} (\bibinfo {year} {2021})}\BibitemShut
  {NoStop}%
\bibitem [{\citenamefont {Mokdad}\ \emph {et~al.}(2019)\citenamefont {Mokdad},
  \citenamefont {Knebel}, \citenamefont {Marin}, \citenamefont {Brison},
  \citenamefont {Phuoc}, \citenamefont {Sopracase}, \citenamefont {Colin},\
  and\ \citenamefont {Braithwaite}}]{mokdad2019structural}%
  \BibitemOpen
  \bibfield  {author} {\bibinfo {author} {\bibfnamefont {J.}~\bibnamefont
  {Mokdad}}, \bibinfo {author} {\bibfnamefont {G.}~\bibnamefont {Knebel}},
  \bibinfo {author} {\bibfnamefont {C.}~\bibnamefont {Marin}}, \bibinfo
  {author} {\bibfnamefont {J.-P.}\ \bibnamefont {Brison}}, \bibinfo {author}
  {\bibfnamefont {V.~T.}\ \bibnamefont {Phuoc}}, \bibinfo {author}
  {\bibfnamefont {R.}~\bibnamefont {Sopracase}}, \bibinfo {author}
  {\bibfnamefont {C.}~\bibnamefont {Colin}},\ and\ \bibinfo {author}
  {\bibfnamefont {D.}~\bibnamefont {Braithwaite}},\ }\bibfield  {title}
  {\bibinfo {title} {Structural, magnetic, and insulator-to-metal transitions
  under pressure in the {GaV$_4$S$_8$} {Mott} insulator: {A} rich phase diagram
  up to 14.7 {GPa}},\ }\href {https://doi.org/10.1103/PhysRevB.100.245101}
  {\bibfield  {journal} {\bibinfo  {journal} {Physical Review B}\ }\textbf
  {\bibinfo {volume} {100}},\ \bibinfo {pages} {245101} (\bibinfo {year}
  {2019})}\BibitemShut {NoStop}%
\bibitem [{\citenamefont {Franke}\ \emph {et~al.}(2018)\citenamefont {Franke},
  \citenamefont {Huddart}, \citenamefont {Hicken}, \citenamefont {Xiao},
  \citenamefont {Blundell}, \citenamefont {Pratt}, \citenamefont {Crisanti},
  \citenamefont {Barker}, \citenamefont {Clark}, \citenamefont
  {{\v{S}}tefan{\v{c}}i{\v{c}}} \emph {et~al.}}]{franke2018magnetic}%
  \BibitemOpen
  \bibfield  {author} {\bibinfo {author} {\bibfnamefont {K.~J.~A.}\
  \bibnamefont {Franke}}, \bibinfo {author} {\bibfnamefont {B.~M.}\
  \bibnamefont {Huddart}}, \bibinfo {author} {\bibfnamefont {T.~J.}\
  \bibnamefont {Hicken}}, \bibinfo {author} {\bibfnamefont {F.}~\bibnamefont
  {Xiao}}, \bibinfo {author} {\bibfnamefont {S.~J.}\ \bibnamefont {Blundell}},
  \bibinfo {author} {\bibfnamefont {F.~L.}\ \bibnamefont {Pratt}}, \bibinfo
  {author} {\bibfnamefont {M.}~\bibnamefont {Crisanti}}, \bibinfo {author}
  {\bibfnamefont {J.~A.~T.}\ \bibnamefont {Barker}}, \bibinfo {author}
  {\bibfnamefont {S.~J.}\ \bibnamefont {Clark}}, \bibinfo {author}
  {\bibfnamefont {A.}~\bibnamefont {{\v{S}}tefan{\v{c}}i{\v{c}}}}, \emph
  {et~al.},\ }\bibfield  {title} {\bibinfo {title} {Magnetic phases of
  skyrmion-hosting {GaV$_4$S$_{8-y}$Se$_y$} ($y$ = 0, 2, 4, 8) probed with muon
  spectroscopy},\ }\href {https://doi.org/10.1103/PhysRevB.98.054428}
  {\bibfield  {journal} {\bibinfo  {journal} {Physical Review B}\ }\textbf
  {\bibinfo {volume} {98}},\ \bibinfo {pages} {054428} (\bibinfo {year}
  {2018})}\BibitemShut {NoStop}%
\bibitem [{\citenamefont {White}\ \emph {et~al.}(2018)\citenamefont {White},
  \citenamefont {Butykai}, \citenamefont {Cubitt}, \citenamefont {Honecker},
  \citenamefont {Dewhurst}, \citenamefont {Kiss}, \citenamefont {Tsurkan},\
  and\ \citenamefont {Bord{\'a}cs}}]{white2018direct}%
  \BibitemOpen
  \bibfield  {author} {\bibinfo {author} {\bibfnamefont {J.~S.}\ \bibnamefont
  {White}}, \bibinfo {author} {\bibfnamefont {A.}~\bibnamefont {Butykai}},
  \bibinfo {author} {\bibfnamefont {R.}~\bibnamefont {Cubitt}}, \bibinfo
  {author} {\bibfnamefont {D.}~\bibnamefont {Honecker}}, \bibinfo {author}
  {\bibfnamefont {C.~D.}\ \bibnamefont {Dewhurst}}, \bibinfo {author}
  {\bibfnamefont {L.~F.}\ \bibnamefont {Kiss}}, \bibinfo {author}
  {\bibfnamefont {V.}~\bibnamefont {Tsurkan}},\ and\ \bibinfo {author}
  {\bibfnamefont {S.}~\bibnamefont {Bord{\'a}cs}},\ }\bibfield  {title}
  {\bibinfo {title} {Direct evidence for cycloidal modulations in the
  thermal-fluctuation-stabilized spin spiral and skyrmion states of
  {Ga}{V}$_4${S}$_8$},\ }\href {https://doi.org/10.1103/PhysRevB.97.020401}
  {\bibfield  {journal} {\bibinfo  {journal} {Physical Review B}\ }\textbf
  {\bibinfo {volume} {97}},\ \bibinfo {pages} {020401(R)} (\bibinfo {year}
  {2018})}\BibitemShut {NoStop}%
\bibitem [{\citenamefont {Clements}\ \emph {et~al.}(2020)\citenamefont
  {Clements}, \citenamefont {Das}, \citenamefont {Pokharel}, \citenamefont
  {Phan}, \citenamefont {Christianson}, \citenamefont {Mandrus}, \citenamefont
  {Prestigiacomo}, \citenamefont {Osofsky},\ and\ \citenamefont
  {Srikanth}}]{clements2020robust}%
  \BibitemOpen
  \bibfield  {author} {\bibinfo {author} {\bibfnamefont {E.~M.}\ \bibnamefont
  {Clements}}, \bibinfo {author} {\bibfnamefont {R.}~\bibnamefont {Das}},
  \bibinfo {author} {\bibfnamefont {G.}~\bibnamefont {Pokharel}}, \bibinfo
  {author} {\bibfnamefont {M.~H.}\ \bibnamefont {Phan}}, \bibinfo {author}
  {\bibfnamefont {A.~D.}\ \bibnamefont {Christianson}}, \bibinfo {author}
  {\bibfnamefont {D.}~\bibnamefont {Mandrus}}, \bibinfo {author} {\bibfnamefont
  {J.~C.}\ \bibnamefont {Prestigiacomo}}, \bibinfo {author} {\bibfnamefont
  {M.~S.}\ \bibnamefont {Osofsky}},\ and\ \bibinfo {author} {\bibfnamefont
  {H.}~\bibnamefont {Srikanth}},\ }\bibfield  {title} {\bibinfo {title} {Robust
  cycloid crossover driven by anisotropy in the skyrmion host {GaV$_4$S$_8$}},\
  }\href {https://doi.org/10.1103/PhysRevB.101.094425} {\bibfield  {journal}
  {\bibinfo  {journal} {Physical Review B}\ }\textbf {\bibinfo {volume}
  {101}},\ \bibinfo {pages} {094425} (\bibinfo {year} {2020})}\BibitemShut
  {NoStop}%
\bibitem [{\citenamefont {Izyumov}(1984)}]{izyumov1984modulated}%
  \BibitemOpen
  \bibfield  {author} {\bibinfo {author} {\bibfnamefont {Y.~A.}\ \bibnamefont
  {Izyumov}},\ }\bibfield  {title} {\bibinfo {title} {Modulated, or
  long-periodic, magnetic structures of crystals},\ }\href
  {https://doi.org/10.1070/pu1984v027n11abeh004120} {\bibfield  {journal}
  {\bibinfo  {journal} {Soviet Physics Uspekhi}\ }\textbf {\bibinfo {volume}
  {27}},\ \bibinfo {pages} {845} (\bibinfo {year} {1984})}\BibitemShut
  {NoStop}%
\bibitem [{\citenamefont {{\v{S}}tefan{\v{c}}i{\v{c}}}\ \emph
  {et~al.}(2020)\citenamefont {{\v{S}}tefan{\v{c}}i{\v{c}}}, \citenamefont
  {Holt}, \citenamefont {Lees}, \citenamefont {Ritter}, \citenamefont
  {Gutmann}, \citenamefont {Lancaster},\ and\ \citenamefont
  {Balakrishnan}}]{stefancic2020establishing}%
  \BibitemOpen
  \bibfield  {author} {\bibinfo {author} {\bibfnamefont {A.}~\bibnamefont
  {{\v{S}}tefan{\v{c}}i{\v{c}}}}, \bibinfo {author} {\bibfnamefont {S.~J.}\
  \bibnamefont {Holt}}, \bibinfo {author} {\bibfnamefont {M.~R.}\ \bibnamefont
  {Lees}}, \bibinfo {author} {\bibfnamefont {C.}~\bibnamefont {Ritter}},
  \bibinfo {author} {\bibfnamefont {M.~J.}\ \bibnamefont {Gutmann}}, \bibinfo
  {author} {\bibfnamefont {T.}~\bibnamefont {Lancaster}},\ and\ \bibinfo
  {author} {\bibfnamefont {G.}~\bibnamefont {Balakrishnan}},\ }\bibfield
  {title} {\bibinfo {title} {Establishing magneto-structural relationships in
  the solid solutions of the skyrmion hosting family of materials:
  {Ga}{V}$_4${S}$_{8-y}${Se}$_y$},\ }\href
  {https://doi.org/10.1038/s41598-020-65676-9} {\bibfield  {journal} {\bibinfo
  {journal} {Scientific Reports}\ }\textbf {\bibinfo {volume} {10}},\ \bibinfo
  {pages} {9813} (\bibinfo {year} {2020})}\BibitemShut {NoStop}%
\bibitem [{\citenamefont {Blundell}(1999)}]{blundell1999spin}%
  \BibitemOpen
  \bibfield  {author} {\bibinfo {author} {\bibfnamefont {S.}~\bibnamefont
  {Blundell}},\ }\bibfield  {title} {\bibinfo {title} {Spin-polarized muons in
  condensed matter physics},\ }\href {https://doi.org/10.1080/001075199181521}
  {\bibfield  {journal} {\bibinfo  {journal} {Contemporary Physics}\ }\textbf
  {\bibinfo {volume} {40}},\ \bibinfo {pages} {175} (\bibinfo {year}
  {1999})}\BibitemShut {NoStop}%
\bibitem [{\citenamefont {Blundell}\ \emph {et~al.}(2021)\citenamefont
  {Blundell}, \citenamefont {De~Renzi}, \citenamefont {Lancaster},\ and\
  \citenamefont {Pratt}}]{blundell2021muon}%
  \BibitemOpen
  \bibfield  {author} {\bibinfo {author} {\bibfnamefont {S.~J.}\ \bibnamefont
  {Blundell}}, \bibinfo {author} {\bibfnamefont {R.}~\bibnamefont {De~Renzi}},
  \bibinfo {author} {\bibfnamefont {T.}~\bibnamefont {Lancaster}},\ and\
  \bibinfo {author} {\bibfnamefont {F.~L.}\ \bibnamefont {Pratt}},\ }\href@noop
  {} {\emph {\bibinfo {title} {Muon Spectroscopy: An Introduction}}}\ (\bibinfo
   {publisher} {Oxford University Press},\ \bibinfo {address} {Oxford},\
  \bibinfo {year} {2021})\BibitemShut {NoStop}%
\bibitem [{sm()}]{sm}%
  \BibitemOpen
  \href@noop {} {\bibinfo {title} {See supplemental material for further
  information on the $\mu^+${SR} technique, and some additional
  data.}}\BibitemShut {Stop}%
\bibitem [{\citenamefont {Khasanov}\ \emph {et~al.}(2016)\citenamefont
  {Khasanov}, \citenamefont {Guguchia}, \citenamefont {Maisuradze},
  \citenamefont {Andreica}, \citenamefont {Elender}, \citenamefont {Raselli},
  \citenamefont {Shermadini}, \citenamefont {Goko}, \citenamefont {Knecht},
  \citenamefont {Morenzoni} \emph {et~al.}}]{khasanov2016high}%
  \BibitemOpen
  \bibfield  {author} {\bibinfo {author} {\bibfnamefont {R.}~\bibnamefont
  {Khasanov}}, \bibinfo {author} {\bibfnamefont {Z.}~\bibnamefont {Guguchia}},
  \bibinfo {author} {\bibfnamefont {A.}~\bibnamefont {Maisuradze}}, \bibinfo
  {author} {\bibfnamefont {D.}~\bibnamefont {Andreica}}, \bibinfo {author}
  {\bibfnamefont {M.}~\bibnamefont {Elender}}, \bibinfo {author} {\bibfnamefont
  {A.}~\bibnamefont {Raselli}}, \bibinfo {author} {\bibfnamefont
  {Z.}~\bibnamefont {Shermadini}}, \bibinfo {author} {\bibfnamefont
  {T.}~\bibnamefont {Goko}}, \bibinfo {author} {\bibfnamefont {F.}~\bibnamefont
  {Knecht}}, \bibinfo {author} {\bibfnamefont {E.}~\bibnamefont {Morenzoni}},
  \emph {et~al.},\ }\bibfield  {title} {\bibinfo {title} {High pressure
  research using muons at the {Paul} {Scherrer} {Institute}},\ }\href
  {https://doi.org/10.1080/08957959.2016.1173690} {\bibfield  {journal}
  {\bibinfo  {journal} {High Pressure Research}\ }\textbf {\bibinfo {volume}
  {36}},\ \bibinfo {pages} {140} (\bibinfo {year} {2016})}\BibitemShut
  {NoStop}%
\bibitem [{\citenamefont {Shermadini}\ \emph {et~al.}(2017)\citenamefont
  {Shermadini}, \citenamefont {Khasanov}, \citenamefont {Elender},
  \citenamefont {Simutis}, \citenamefont {Guguchia}, \citenamefont {Kamenev},\
  and\ \citenamefont {Amato}}]{shermadini2017low}%
  \BibitemOpen
  \bibfield  {author} {\bibinfo {author} {\bibfnamefont {Z.}~\bibnamefont
  {Shermadini}}, \bibinfo {author} {\bibfnamefont {R.}~\bibnamefont
  {Khasanov}}, \bibinfo {author} {\bibfnamefont {M.}~\bibnamefont {Elender}},
  \bibinfo {author} {\bibfnamefont {G.}~\bibnamefont {Simutis}}, \bibinfo
  {author} {\bibfnamefont {Z.}~\bibnamefont {Guguchia}}, \bibinfo {author}
  {\bibfnamefont {K.~V.}\ \bibnamefont {Kamenev}},\ and\ \bibinfo {author}
  {\bibfnamefont {A.}~\bibnamefont {Amato}},\ }\bibfield  {title} {\bibinfo
  {title} {A low-background piston--cylinder-type hybrid high pressure cell for
  muon-spin rotation/relaxation experiments},\ }\href
  {https://doi.org/10.1080/08957959.2017.1373773} {\bibfield  {journal}
  {\bibinfo  {journal} {High Pressure Research}\ }\textbf {\bibinfo {volume}
  {37}},\ \bibinfo {pages} {449} (\bibinfo {year} {2017})}\BibitemShut
  {NoStop}%
\bibitem [{\citenamefont {Pratt}(2000)}]{pratt2000wimda}%
  \BibitemOpen
  \bibfield  {author} {\bibinfo {author} {\bibfnamefont {F.~L.}\ \bibnamefont
  {Pratt}},\ }\bibfield  {title} {\bibinfo {title} {{WiMDA}: a muon data
  analysis program for the {Windows} {PC}},\ }\href
  {https://doi.org/10.1016/S0921-4526(00)00328-8} {\bibfield  {journal}
  {\bibinfo  {journal} {Physica B: Condensed Matter}\ }\textbf {\bibinfo
  {volume} {289}},\ \bibinfo {pages} {710} (\bibinfo {year}
  {2000})}\BibitemShut {NoStop}%
\bibitem [{\citenamefont {James}\ and\ \citenamefont
  {Roos}(1975)}]{james1975minuit}%
  \BibitemOpen
  \bibfield  {author} {\bibinfo {author} {\bibfnamefont {F.}~\bibnamefont
  {James}}\ and\ \bibinfo {author} {\bibfnamefont {M.}~\bibnamefont {Roos}},\
  }\bibfield  {title} {\bibinfo {title} {{MINUIT}: a system for function
  minimization and analysis of the parameter errors and corrections},\ }\href
  {https://doi.org/10.1016/0010-4655(75)90039-9} {\bibfield  {journal}
  {\bibinfo  {journal} {Computer Physiscs Commununications}\ }\textbf {\bibinfo
  {volume} {10}},\ \bibinfo {pages} {343} (\bibinfo {year} {1975})}\BibitemShut
  {NoStop}%
\bibitem [{\citenamefont {iminuit team}(2021)}]{iminuit}%
  \BibitemOpen
  \bibfield  {author} {\bibinfo {author} {\bibnamefont {iminuit team}},\
  }\href@noop {} {\bibinfo {title} {iminuit -- a python interface to minuit}},\
  \bibinfo {howpublished} {\url{https://github.com/scikit-hep/iminuit}}
  (\bibinfo {year} {Accessed: 16-09-2021})\BibitemShut {NoStop}%
\bibitem [{\citenamefont {Bonf{\`a}}\ \emph {et~al.}(2018)\citenamefont
  {Bonf{\`a}}, \citenamefont {Onuorah},\ and\ \citenamefont
  {De~Renzi}}]{bonfa2018introduction}%
  \BibitemOpen
  \bibfield  {author} {\bibinfo {author} {\bibfnamefont {P.}~\bibnamefont
  {Bonf{\`a}}}, \bibinfo {author} {\bibfnamefont {I.~J.}\ \bibnamefont
  {Onuorah}},\ and\ \bibinfo {author} {\bibfnamefont {R.}~\bibnamefont
  {De~Renzi}},\ }\bibfield  {title} {\bibinfo {title} {Introduction and a quick
  look at {MUESR}, the {M}agnetic structure and m{U}on {E}mbedding {S}ite
  {R}efinement suite},\ }in\ \href {https://doi.org/10.7566/JPSCP.21.011052}
  {\emph {\bibinfo {booktitle} {Proceedings of the 14th International
  Conference on Muon Spin Rotation, Relaxation and Resonance ($\mu$SR2017)}}}\
  (\bibinfo {year} {2018})\ p.\ \bibinfo {pages} {011052}\BibitemShut {NoStop}%
\bibitem [{\citenamefont {Beg}\ \emph {et~al.}(2017)\citenamefont {Beg},
  \citenamefont {Pepper},\ and\ \citenamefont {Fangohr}}]{beg2017user}%
  \BibitemOpen
  \bibfield  {author} {\bibinfo {author} {\bibfnamefont {M.}~\bibnamefont
  {Beg}}, \bibinfo {author} {\bibfnamefont {R.~A.}\ \bibnamefont {Pepper}},\
  and\ \bibinfo {author} {\bibfnamefont {H.}~\bibnamefont {Fangohr}},\
  }\bibfield  {title} {\bibinfo {title} {User interfaces for computational
  science: {A} domain specific language for {OOMMF} embedded in {Python}},\
  }\href {https://doi.org/10.1063/1.4977225} {\bibfield  {journal} {\bibinfo
  {journal} {AIP Advances}\ }\textbf {\bibinfo {volume} {7}},\ \bibinfo {pages}
  {056025} (\bibinfo {year} {2017})}\BibitemShut {NoStop}%
\bibitem [{\citenamefont {Beg}\ \emph {et~al.}(2021)\citenamefont {Beg},
  \citenamefont {Pepper}, \citenamefont {Kluyver}, \citenamefont {Mulkers},
  \citenamefont {Leliaert},\ and\ \citenamefont {Fangohr}}]{ubermag}%
  \BibitemOpen
  \bibfield  {author} {\bibinfo {author} {\bibfnamefont {M.}~\bibnamefont
  {Beg}}, \bibinfo {author} {\bibfnamefont {R.~A.}\ \bibnamefont {Pepper}},
  \bibinfo {author} {\bibfnamefont {T.}~\bibnamefont {Kluyver}}, \bibinfo
  {author} {\bibfnamefont {J.}~\bibnamefont {Mulkers}}, \bibinfo {author}
  {\bibfnamefont {J.}~\bibnamefont {Leliaert}},\ and\ \bibinfo {author}
  {\bibfnamefont {H.}~\bibnamefont {Fangohr}},\ }\href
  {https://doi.org/10.5281/zenodo.3539496} {\bibinfo {title} {ubermag: Meta
  package for ubermag project.}} (\bibinfo {year} {Accessed:
  16-09-2021})\BibitemShut {NoStop}%
\bibitem [{\citenamefont {Padmanabhan}\ \emph {et~al.}(2019)\citenamefont
  {Padmanabhan}, \citenamefont {Sekiguchi}, \citenamefont {Versteeg},
  \citenamefont {Slivina}, \citenamefont {Tsurkan}, \citenamefont
  {Bord{\'a}cs}, \citenamefont {K{\'e}zsm{\'a}rki},\ and\ \citenamefont
  {Van~Loosdrecht}}]{padmanabhan2019optically}%
  \BibitemOpen
  \bibfield  {author} {\bibinfo {author} {\bibfnamefont {P.}~\bibnamefont
  {Padmanabhan}}, \bibinfo {author} {\bibfnamefont {F.}~\bibnamefont
  {Sekiguchi}}, \bibinfo {author} {\bibfnamefont {R.~B.}\ \bibnamefont
  {Versteeg}}, \bibinfo {author} {\bibfnamefont {E.}~\bibnamefont {Slivina}},
  \bibinfo {author} {\bibfnamefont {V.}~\bibnamefont {Tsurkan}}, \bibinfo
  {author} {\bibfnamefont {S.}~\bibnamefont {Bord{\'a}cs}}, \bibinfo {author}
  {\bibfnamefont {I.}~\bibnamefont {K{\'e}zsm{\'a}rki}},\ and\ \bibinfo
  {author} {\bibfnamefont {P.~H.~M.}\ \bibnamefont {Van~Loosdrecht}},\
  }\bibfield  {title} {\bibinfo {title} {Optically driven collective spin
  excitations and magnetization dynamics in the {N}{\'e}el-type skyrmion host
  {GaV$_4$S$_8$}},\ }\href {https://doi.org/10.1103/PhysRevLett.122.107203}
  {\bibfield  {journal} {\bibinfo  {journal} {Physical Review Letters}\
  }\textbf {\bibinfo {volume} {122}},\ \bibinfo {pages} {107203} (\bibinfo
  {year} {2019})}\BibitemShut {NoStop}%
\bibitem [{\citenamefont {Routledge}\ \emph {et~al.}(2021)\citenamefont
  {Routledge}, \citenamefont {Vir}, \citenamefont {Cook}, \citenamefont
  {Murgatroyd}, \citenamefont {Ahmed}, \citenamefont {Savvin}, \citenamefont
  {Claridge},\ and\ \citenamefont {Alaria}}]{routledge2021mode}%
  \BibitemOpen
  \bibfield  {author} {\bibinfo {author} {\bibfnamefont {K.}~\bibnamefont
  {Routledge}}, \bibinfo {author} {\bibfnamefont {P.}~\bibnamefont {Vir}},
  \bibinfo {author} {\bibfnamefont {N.}~\bibnamefont {Cook}}, \bibinfo {author}
  {\bibfnamefont {P.~A.~E.}\ \bibnamefont {Murgatroyd}}, \bibinfo {author}
  {\bibfnamefont {S.~J.}\ \bibnamefont {Ahmed}}, \bibinfo {author}
  {\bibfnamefont {S.~N.}\ \bibnamefont {Savvin}}, \bibinfo {author}
  {\bibfnamefont {J.~B.}\ \bibnamefont {Claridge}},\ and\ \bibinfo {author}
  {\bibfnamefont {J.}~\bibnamefont {Alaria}},\ }\bibfield  {title} {\bibinfo
  {title} {Mode crystallography analysis through the structural phase
  transition and magnetic critical behavior of the lacunar spinel
  {GaMo$_4$Se$_8$}},\ }\href {https://doi.org/10.1021/acs.chemmater.1c01448}
  {\bibfield  {journal} {\bibinfo  {journal} {Chemistry of Materials}\ }\textbf
  {\bibinfo {volume} {33}},\ \bibinfo {pages} {5718} (\bibinfo {year}
  {2021})}\BibitemShut {NoStop}%
\bibitem [{\citenamefont {Schueller}\ \emph {et~al.}(2020)\citenamefont
  {Schueller}, \citenamefont {Kitchaev}, \citenamefont {Zuo}, \citenamefont
  {Bocarsly}, \citenamefont {Cooley}, \citenamefont {Van~der Ven},
  \citenamefont {Wilson},\ and\ \citenamefont
  {Seshadri}}]{schueller2020structural}%
  \BibitemOpen
  \bibfield  {author} {\bibinfo {author} {\bibfnamefont {E.~C.}\ \bibnamefont
  {Schueller}}, \bibinfo {author} {\bibfnamefont {D.~A.}\ \bibnamefont
  {Kitchaev}}, \bibinfo {author} {\bibfnamefont {J.~L.}\ \bibnamefont {Zuo}},
  \bibinfo {author} {\bibfnamefont {J.~D.}\ \bibnamefont {Bocarsly}}, \bibinfo
  {author} {\bibfnamefont {J.~A.}\ \bibnamefont {Cooley}}, \bibinfo {author}
  {\bibfnamefont {A.}~\bibnamefont {Van~der Ven}}, \bibinfo {author}
  {\bibfnamefont {S.~D.}\ \bibnamefont {Wilson}},\ and\ \bibinfo {author}
  {\bibfnamefont {R.}~\bibnamefont {Seshadri}},\ }\bibfield  {title} {\bibinfo
  {title} {Structural evolution and skyrmionic phase diagram of the lacunar
  spinel {GaMo$_4$Se$_8$}},\ }\href
  {https://doi.org/10.1103/PhysRevMaterials.4.064402} {\bibfield  {journal}
  {\bibinfo  {journal} {Physical Review Materials}\ }\textbf {\bibinfo {volume}
  {4}},\ \bibinfo {pages} {064402} (\bibinfo {year} {2020})}\BibitemShut
  {NoStop}%
\bibitem [{\citenamefont {M{\"u}ller}\ \emph {et~al.}(2006)\citenamefont
  {M{\"u}ller}, \citenamefont {Kockelmann},\ and\ \citenamefont
  {Johrendt}}]{muller2006magnetic}%
  \BibitemOpen
  \bibfield  {author} {\bibinfo {author} {\bibfnamefont {H.}~\bibnamefont
  {M{\"u}ller}}, \bibinfo {author} {\bibfnamefont {W.}~\bibnamefont
  {Kockelmann}},\ and\ \bibinfo {author} {\bibfnamefont {D.}~\bibnamefont
  {Johrendt}},\ }\bibfield  {title} {\bibinfo {title} {The magnetic structure
  and electronic ground states of {Mott} insulators {GeV$_4$S$_8$} and
  {GaV$_4$S$_8$}},\ }\href {https://doi.org/10.1021/cm052809m} {\bibfield
  {journal} {\bibinfo  {journal} {Chemistry of Materials}\ }\textbf {\bibinfo
  {volume} {18}},\ \bibinfo {pages} {2174} (\bibinfo {year}
  {2006})}\BibitemShut {NoStop}%
\bibitem [{\citenamefont {Bichler}\ \emph {et~al.}(2008)\citenamefont
  {Bichler}, \citenamefont {Zinth}, \citenamefont {Johrendt}, \citenamefont
  {Heyer}, \citenamefont {Forthaus}, \citenamefont {Lorenz},\ and\
  \citenamefont {Abd-Elmeguid}}]{bichler2008structural}%
  \BibitemOpen
  \bibfield  {author} {\bibinfo {author} {\bibfnamefont {D.}~\bibnamefont
  {Bichler}}, \bibinfo {author} {\bibfnamefont {V.}~\bibnamefont {Zinth}},
  \bibinfo {author} {\bibfnamefont {D.}~\bibnamefont {Johrendt}}, \bibinfo
  {author} {\bibfnamefont {O.}~\bibnamefont {Heyer}}, \bibinfo {author}
  {\bibfnamefont {M.~K.}\ \bibnamefont {Forthaus}}, \bibinfo {author}
  {\bibfnamefont {T.}~\bibnamefont {Lorenz}},\ and\ \bibinfo {author}
  {\bibfnamefont {M.~M.}\ \bibnamefont {Abd-Elmeguid}},\ }\bibfield  {title}
  {\bibinfo {title} {Structural and magnetic phase transitions of the v
  4-cluster compound {GeV$_4$S$_8$}},\ }\href
  {https://doi.org/10.1103/PhysRevB.77.212102} {\bibfield  {journal} {\bibinfo
  {journal} {Physical Review B}\ }\textbf {\bibinfo {volume} {77}},\ \bibinfo
  {pages} {212102} (\bibinfo {year} {2008})}\BibitemShut {NoStop}%
\bibitem [{\citenamefont {Ruff}\ \emph {et~al.}(2015)\citenamefont {Ruff},
  \citenamefont {Widmann}, \citenamefont {Lunkenheimer}, \citenamefont
  {Tsurkan}, \citenamefont {Bord{\'a}cs}, \citenamefont {K{\'e}zsm{\'a}rki},\
  and\ \citenamefont {Loidl}}]{ruff2015multiferroicity}%
  \BibitemOpen
  \bibfield  {author} {\bibinfo {author} {\bibfnamefont {E.}~\bibnamefont
  {Ruff}}, \bibinfo {author} {\bibfnamefont {S.}~\bibnamefont {Widmann}},
  \bibinfo {author} {\bibfnamefont {P.}~\bibnamefont {Lunkenheimer}}, \bibinfo
  {author} {\bibfnamefont {V.}~\bibnamefont {Tsurkan}}, \bibinfo {author}
  {\bibfnamefont {S.}~\bibnamefont {Bord{\'a}cs}}, \bibinfo {author}
  {\bibfnamefont {I.}~\bibnamefont {K{\'e}zsm{\'a}rki}},\ and\ \bibinfo
  {author} {\bibfnamefont {A.}~\bibnamefont {Loidl}},\ }\bibfield  {title}
  {\bibinfo {title} {Multiferroicity and skyrmions carrying electric
  polarization in {GaV$_4$S$_8$}},\ }\href
  {https://doi.org/10.1126/sciadv.1500916} {\bibfield  {journal} {\bibinfo
  {journal} {Science advances}\ }\textbf {\bibinfo {volume} {1}},\ \bibinfo
  {pages} {e1500916} (\bibinfo {year} {2015})}\BibitemShut {NoStop}%
\bibitem [{\citenamefont {Fujima}\ \emph {et~al.}(2017)\citenamefont {Fujima},
  \citenamefont {Abe}, \citenamefont {Tokunaga},\ and\ \citenamefont
  {Arima}}]{fujima2017thermodynamically}%
  \BibitemOpen
  \bibfield  {author} {\bibinfo {author} {\bibfnamefont {Y.}~\bibnamefont
  {Fujima}}, \bibinfo {author} {\bibfnamefont {N.}~\bibnamefont {Abe}},
  \bibinfo {author} {\bibfnamefont {Y.}~\bibnamefont {Tokunaga}},\ and\
  \bibinfo {author} {\bibfnamefont {T.}~\bibnamefont {Arima}},\ }\bibfield
  {title} {\bibinfo {title} {Thermodynamically stable skyrmion lattice at low
  temperatures in a bulk crystal of lacunar spinel {GaV$_4$Se$_8$}},\ }\href
  {https://doi.org/10.1103/PhysRevB.95.180410} {\bibfield  {journal} {\bibinfo
  {journal} {Physical Review B}\ }\textbf {\bibinfo {volume} {95}},\ \bibinfo
  {pages} {180410} (\bibinfo {year} {2017})}\BibitemShut {NoStop}%
\end{thebibliography}%
	
\end{document}